\documentclass[a4paper, 10pt]{article}
\usepackage[utf8]{inputenc}
\usepackage{fancyhdr}
\usepackage{graphicx}
\usepackage{mathrsfs}
\usepackage{amsmath}
\usepackage{tikz}
\usepackage{hyperref}
\usepackage{amssymb}
\usepackage{siunitx}
\usepackage{caption}
\usepackage{subcaption}
\usepackage[style=numeric,backend=bibtex]{biblatex}
\usepackage[outdir=./]{epstopdf}
\usepackage{multicol}
\newcommand\scalemath[2]{\scalebox{#1}{\mbox{\ensuremath{\displaystyle #2}}}}
\begin{document}
\title{Prediction of dynamical systems using geometric constraints imposed by observations}
\author{Saurabh Dixit\footnote{Email: saurabhdixit@iisc.ac.in}  \qquad
Soumyendu Raha\footnote{Email: raha@iisc.ac.in} \\ \small{\textit{Department of Computational and Data Sciences,}} \\ \small{\textit{Indian Institute of Science, Bangalore, India, 560012}}}
\date{}
\lhead{S Dixit and S Raha}
\maketitle
\begin{abstract}
Solution of Ordinary Differential Equation(ODE) model of dynamical system may not agree with its observed values. 
Often this discrepancy can be attributed to unmodeled forcings in the evolution rule of the dynamical system. 
In this article, an approach for data-based model improvement is described which exploits the geometric constraint imposed by the system observations to estimate these unmodeled terms. The nominal model is augmented using these extra forcing terms to make predictions. This approach is applied to navigational satellite orbit prediction to bring down the error to $\approx 12\%$ of the error when using the nominal force model for a 2-hour prediction. In another example improved temperature predictions over the nominal heat equation are obtained for one-dimensional conduction.  
\end{abstract}
\section{Introduction}
Classical approach for modeling time-evolving systems involves formulating ODEs governing its evolution using fundamental physical principles. These ODEs are solved forward in time to obtain future states of the system. In practice, the governing dynamics of a system is attributed to concurrence of more than one physical phenomenon.
Unmodeled dynamics and inaccurate parameters in such models result in prediction errors. There are other modeling approaches
which build models solely based on historical data. Such ``Black-box'' model is a product of the data with which
it is built and quality of model depends on the quality of data. It also disregards any change in the applicability of the dynamics or its 
parameters between ``training'' and prediction time interval. This suggests us that a combination of the two approaches can be used
to overcome the inadequency of either approaches to make useful predictions for real-world problems.\par 
Many physical laws can be formulated as a problem of finding extrema of a functional using variational principles. 
Such formalism can incorporate constraints on the system variables which leads to the Euler equations. These equations can be referenced in standard texts on classical mechanics such as \cite{Goldstein,Morse}. 
The Euler equations with suitable discretization along with the constraint equations give rise to a system of Differential-Algebraic Equations (DAE).
Numerical schemes for solving the DAEs of the forms that most often occur in practice can be found in \cite{Petzold}. \par
In the present work, we combine the two modeling approaches by considering system observations as constraints on its dynamics. This leads to additional terms involving Lagrange multipliers in the evolution rule for constrained dynamical system. The DAE occuring with such formulation is solved for the Lagrange multipliers giving us additional constraint forces. With suitable analysis of the Lagrange multiplier data thus obtained, we augment the evolution rule with additional forcing terms for prediction.\par
The outline of this paper is as follows:
Section 2 discusses the mathematical formulation of the approach and the relevant background.
In section 3, the formulation of section 2 is tailored for the problem of satellite orbit prediction. It is then used to predict orbit of a geostationary satellite of BeiDou Navigation Satellite System (BDS).
In section 4, this method is applied to another example of temperature prediction in one-dimensional conduction. DAE formulation for heat equation with constraints along with application using data of different nature is the main highlight of this section. We summarize our results and conclude the findings of this article in section 5.
\section{Background and motivation}
Evolution equations for a constrained dynamical system with known Lagrange density can be obtained from the Euler equations \cite{Morse}. 
Let $\boldsymbol{\omega}=(\omega_1\dots \omega_m)$ be the independent parameters of the system, 
$\mathbf{h}=(h_1,\dots h_n)$ be the system variables as a function  of $\boldsymbol{\omega}$. Let
$L(\mathbf{h},\frac{\partial \mathbf{h}}{\partial \boldsymbol{\omega}},\boldsymbol{\omega})$ be the Lagrange density function for the given 
dynamical system subject to k constraints given by:
\begin{equation}
 \mathbf{M}\Big(\mathbf{h},\frac{\partial\mathbf{h}}{\partial \boldsymbol{\omega}},\boldsymbol{\omega}\Big)=\mathbf{0}, \label{1}
\end{equation}
Denoting $\frac{\partial h_i}{\partial \omega_j}$ by $h_{ij}$, the Euler equations are given as:
\begin{equation}
 \sum_{j=1}^{m}\frac{\partial}{\partial \omega_j} \frac{\partial L^{'}}{\partial h_{ij}}=\frac{\partial L{'}}{\partial h_i}, \label{2}
\end{equation}
where 
\begin{align}
 L'=L+ \sum_{l=1}^{k}\lambda_l(\boldsymbol{\omega}) M_l.  \label{3}
\end{align}
In equation \ref{3} $\lambda_l$ is the Lagrange multiplier corresponding to the $l^{th}$ constraint denoted by $M_l$.
Equation \ref{2} represents set of $n$ equations with i taking values from 1 to n. In many problems of interest, equations (\ref{2}) can be discretized
to give us a system of ODEs. Along with the algebraic constraints (\ref{1}), it results in a DAE system.\par
We consider that some functions of system variables are measured and are available as observation data. The observed variables, being functions of the system variables, lead to constraint equation of the form \ref{1} on the system dynamics. Additional forces experienced by the system to constrain the dynamics on the manifold formed by the
observational constraint equations are computed in terms of Lagrange multipliers. If the observations are noise-free, these constraint forces can be computed accurately. These Lagrange multipliers are stored as 
functions of $\boldsymbol{\omega}$ and $\mathbf{h}$. For state prediction, equation of the form (\ref{2}) is used by substituting the 
Lagrange multipliers computed from the historical data either directly or by suitable model fitting between the Lagrange multipliers
and the variables $\boldsymbol{\omega}$ and $\mathbf{h}$.\par
This idea is applied to improve satellite orbit prediction and temperature prediction for one-dimensional heat conduction experiment using observed data for these systems. Implementation of this method to these problems is discussed in detail in the following sections.

\section{Satellite Orbit Prediction}
In this section, the approach discussed in the last section is elaborated and applied for improving Global Navigation Satellite System(GNSS) satellite orbit prediction. In the following paragraphs, force equation of a satellite in the gravitational field of earth subjected to constraints is derived. These equations are used to compute extra forces required to satisfy the constraints due to observations.\par
Satellite orbiting around the earth experiences not only the gravitational pull from the earth but also other perturbation forces including the effect due to the oblateness of the earth, solar radiation pressure, gravity of moon and sun and other celestial bodies ~\cite{Misra}. We consider the reduced dynamics model--with only the gravitational force due to the point-mass earth--as the nominal force model for the satellite.\par
Euler equations \ref{2} are used to determine the equation of motion of conservative system in classical mechanics using the Hamilton's principle.
The independent parameter $\boldsymbol{\omega}$ in this case is real-valued time t, the dependent variables are the positions
in configuration space in generalised coordinates and Lagrange density function is the difference between Kinetic and Potential energies ~\cite{Morse}. 
The gravitational potential V of a satellite of mass $m$ at a distance $r$ from the earth, with both earth and the satellite assumed to be point masses, is given by,
\begin{equation}
 V({r})=-\frac{GMm}{r} \label{4},
\end{equation}
where M is the mass of the earth.
Let $\mathbf{r}=(x_1,x_2,x_3)^T$ be the coordinates of the satellite in the International Celestial Reference Frame (ICRF) and let $\boldsymbol v$ be its velocity with components $v_1$, $v_2$ and $v_3$. Then the Kinetic
energy T of the satellite is given by,
\begin{equation}
 T=\frac{1}{2}m(v_1^2+v_2^2+v_3^2) \label{5}.
\end{equation}

The Lagrangian L is given by $(T-V)$. If the system is subjected to k holonomic or semiholonomic constraints, then from equation (\ref{3})  we get,
\begin{equation}
 L^{'}= \frac{1}{2}m(v_1^2+v_2^2+v_3^2)+\frac{GMm}{r}+\sum_{l=1}^{k}\mu_l(t) M_l \label{6},
\end{equation}
where $\mu_l(t)$ denotes the $l^{th}$ Lagrange multiplier. Flannery ~\cite{Flannery} has shown that the Lagrangian density $L^{'}$ can be directly substituted in (\ref{2}) for
semiholonomic constraints.
Substituting these values in equation (\ref{2}), for $i \in \{1,2,3\}$ we have,
\begin{equation}
\frac{d(mv_i)}{dt}+\frac{d}{dt}(\sum_{l=1}^{k}\mu_l(t)\frac{\partial M_l}{\partial v_i})=-\frac{GMmx_i}{(x_1^2+x_2^2+x_3^2)^{\frac{3}{2}}}+\sum_{l=1}^{k}\mu_l(t)\frac{\partial M_l}{\partial x_i}. 
\label{7}
\end{equation}
In particular, if $k=3$ and the constraint $M_l$ is of the form:
\begin{equation}
 M_l:=v_l-\phi_{l}(t)=0 \label{8},
\end{equation}
Then (\ref{7}) becomes,
\begin{equation}
 \frac{d(mv_i)}{dt}+\frac{d}{dt}(\mu_i(t))=-\frac{GMmx_i}{(x_1^2+x_2^2+x_3^2)^{\frac{3}{2}}} \label{9}.
\end{equation}
It is to be noted that equation \ref{8} can be integrated to obtain the position coordinate $x_i$, and can be expressed as an equivalent constraint on the position of the satellite of the form:
\begin{align}
M^p_l:=x_i-\phi^p_{l}(t)=0. \label{10}
\end{align}
If $M_l$ is replaced by $M^p_l$ in \ref{7} and solved with the constraints (\ref{10}), another set of Lagrange multipliers are obtained,
say $\bar{\lambda}_i(t)$. The relationship between $\bar{\lambda}_i(t)$ and $\mu_i(t)$ is  $\bar{\lambda}_i(t) =- \frac{d}{dt}\mu_i(t)$ as noted in
\cite{Flannery}. Let $\lambda_i(t)=\frac{\bar{\lambda}_i(t)}{m}$. Dividing (\ref{9}) by $m$ and writing in terms of $\lambda_i(t)$ we get,
\begin{equation}
 \frac{dv_i}{dt}=-\frac{GMx_i}{(x_1^2+x_2^2+x_3^2)^{\frac{3}{2}}}+\lambda_i \label{11}.
\end{equation}
The nominal force model without the constraints can be obtained by simply using $T-V$ as the Lagrangian density to obtain
the Newton's Law of gravitation and the equation for the nominal acceleration is simply given by:
\begin{equation}
  \frac{dv_i}{dt}=-\frac{GMx_i}{(x_1^2+x_2^2+x_3^2)^{\frac{3}{2}}} \label{12}.
\end{equation}
The jacobian $\frac{\partial \boldsymbol M}{\partial \boldsymbol v}$ is identity for constraint \ref{8} and thus the additional acceleration that appears in the nominal model to satisfy these constraint equations is just  $\lambda_i$. Equations (\ref{11}) and (\ref{8}) for all i's form a DAE system of index-2  along the solution $(\mathbf{x}(t),\boldsymbol{\lambda}(t))$[\cite{Petzold}], where $\boldsymbol{\lambda}=(\lambda_1, \lambda_3, \lambda_3)^T$. It is solved for $\boldsymbol{\lambda}$ to
obtain the extra acceleration needed to follow the data constraints.  The value of $\phi_i(t)$ in equation (\ref{8}) is determined by historical accurate velocity data for discrete points in time. The DAE solution lets us obtain samples of ``additional forces''
acting on the satellite for different sectors of space from the historical data.\par
International GNSS Service(IGS) provides GNSS satellite coordinates with respect to the International Terrestial Reference Frame(ITRF) in SP3 format \cite{hilla2016}. These satellite coordinates are available every 15 minutes and have an RMS accuracy of $\approx 2.5$ cm \cite{Montenbruck1,Montenbruck2,johnston}.
The coordinates are transformed to the International Celestial Reference Frame(ICRF) using the earth orientation matrices obtained from \cite{EOP}. 
The satellite positions thus obtained in ICRF have a frequency of 4 data points per hour. This positional data is interpolated to a frequency of 1 data point per second to facilitate the computation of finite difference approximation of the velocities. A scheme for interpolation of satellite coordinates with millimeter level rms accuracy is described in \cite{Horemuz}.\par
 Following the work of \cite{Horemuz}, a 16 degree polynomial is fitted to four hour positional data. The coefficients of the interpolating polynomial are used to generate the interpolated positional data every second for the central two hours i.e the second and the third hour. Again a new polynomial is fitted for
four hourly data starting from the third hour in a moving window fashion and the interpolated polynomial coefficients are used
to generate data for the fourth and fifth hour, and so on. Let $[x_{1m}(t),x_{2m}(t),x_{3m}(t)]^T$ denote the interpolated position vector at time t.
$i^{th}$ component  of velocity at each epoch is computed using a finite difference approximation of the interpolated positions as follows,
\begin{equation}
 v_{im}(t)=\frac{x_{im}(t+\Delta t)-x_{im}(t)}{\Delta t} \label{13},
\end{equation}
$\Delta t=1$s since interpolated position coordinates are available every second. Velocities thus obtained at various time points form the \textit{observation dataset}.\par
The velocity observations from the dataset can directly be substituted in algebraic equation \ref{8} to obtain the values of $\phi_i(t)$ at discrete time epochs.
The DAE system \ref{11} and \ref{8} is discretized numerically using Trapezoidal method ~\cite{Petzold,Raha} with a time-step $h=\Delta t =1$s and solved for $(\mathbf{x}(t),\boldsymbol{\lambda}(t))$ by substituting the known values of $v_{im}(t)$ for $\phi_i(t)$ at discrete time points. 
The details of the numerical scheme are as follows: Denote time $t^k = kh$, where k is the non-negative integer step number and $h$ is the step size. Let $x^k_i$, $v^k_i$, $v^k_{im}$ and $\lambda^k_i$
denote the value of the $i^{th}$ component of position, velocity, observed velocity and Lagrange multiplier at the $k^{th}$ time-step.
The discretization for the $i^{th}$ $(i\in\{1,2,3\})$ component of the DAE system \ref{11} and \ref{8} is given as :
\\ 
\begin{align}
 x_i^{k+1}&=x_i^k+hv_i^k \label{sc1} \\
 v_i^{k+1}&=v_i^{k}+\frac{1}{2}h (p_i^{k}+ p_i^{k+1}) \label{sc2} \\
 p_i^{k+1}&=\frac{-GMx_i^{k+1}}{[({x_1^{k+1}})^2+({x_2^{k+1}})^2+({x_3^{k+1}})^2]^\frac{3}{2}}+\lambda_i^{k+1} \label{sc3} \\
 v_i^{k+1}&=v_{im}^{k+1} \label{16}
\end{align}
Initial time-step for numerical solution is at $t^1$ instead of $t^0$, as the finite difference approximation of initial
acceralation can only be made at $t^1$, if the first available data point is at $t^0$.
To get consistent initial conditions $p_i^{1}$ is initialized with the approximate observed acceleration at first time-step, 
given as:
\begin{align}
 p_i^{1}&=\frac{x_{im}^0-2x_{im}^1+x_{im}^2}{{\Delta t}^2} \label{20}
\end{align}
Other variables are intialized as:
$x_i^1=x_{im}^1$, $v_i^1=v_{im}^1$, $\lambda_i^1=0$.  
\par
In this way, numerical solution of this DAE using the observed data gives the values of $\boldsymbol{\lambda}=[\lambda_1,\lambda_3,\lambda_3]^T$ 
as a function of time and satellite position.These $\boldsymbol{\lambda}$ values are stored in a dataset against the position 
vector $(x_1,x_2,x_3)^T=\boldsymbol{r}$ of the satellite in ICRF for each 
measurement epoch. Such a dataset is generated for few days using the SP3 positional data of the satellite. Henceforth, we call this dataset the $\lambda$-dataset.\par
 To predict the state of the satellite (position and velocities) at a future time, ODE (\ref{11}) is solved numerically using the trapezoidal method to give $\mathbf{r}(t)$ for time t.
The intial conditions are given by ,
${x}_i^0=x_{mi}^0$, ${x}_i^1=x_{mi}^1$ and $v_i^0=\frac{1}{\Delta t}({x}_i^1-{x}_i^0)$.
The value of $\boldsymbol{\lambda}$ is chosen from the $\lambda$-dataset at time t by selecting the $\mathbf{\boldsymbol\lambda}$ corresponding to the
$\mathbf{r_i}$ such that $\|\mathbf{r}(t)-\mathbf{r_i}\| \leq \|\mathbf{r}(t)-\mathbf{r_j}\|$ for all $\mathbf{r_j}$ in the $\lambda$-dataset.\par
\begin{figure}
\centering
		\includegraphics[width=1\linewidth]{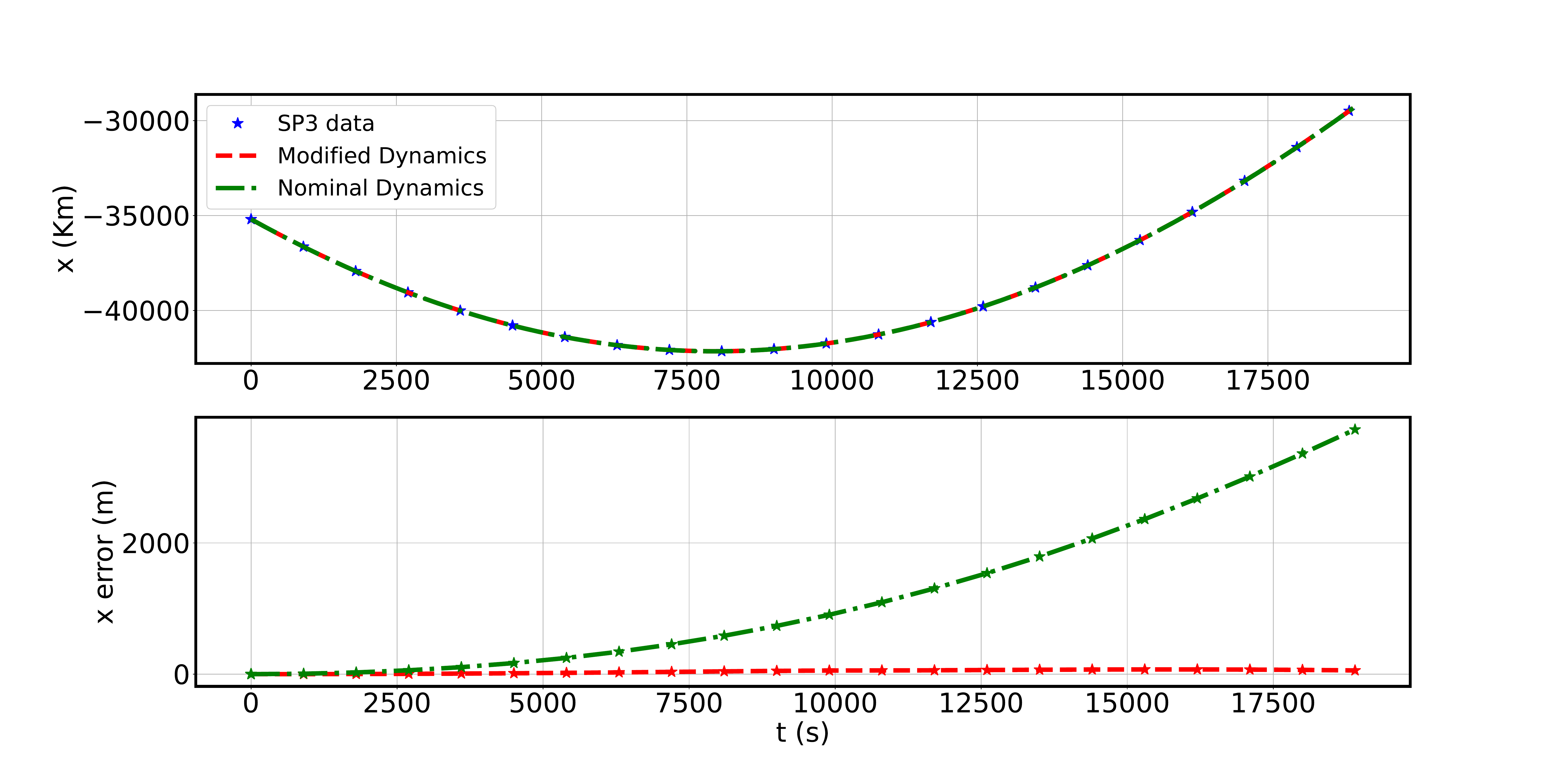}
\centering
 		\includegraphics[width=1\linewidth]{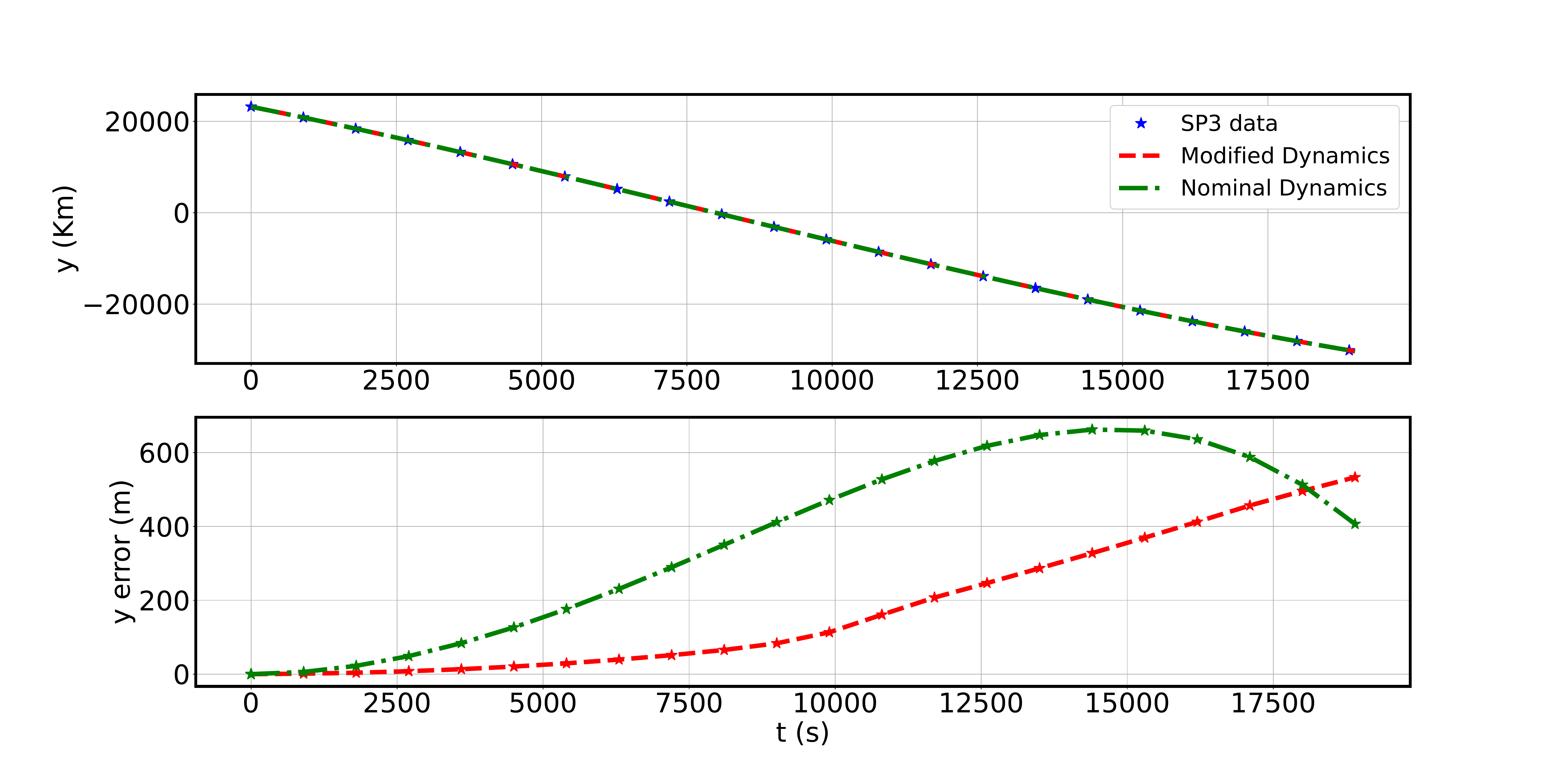}
\centering
		\includegraphics[width=1\linewidth]{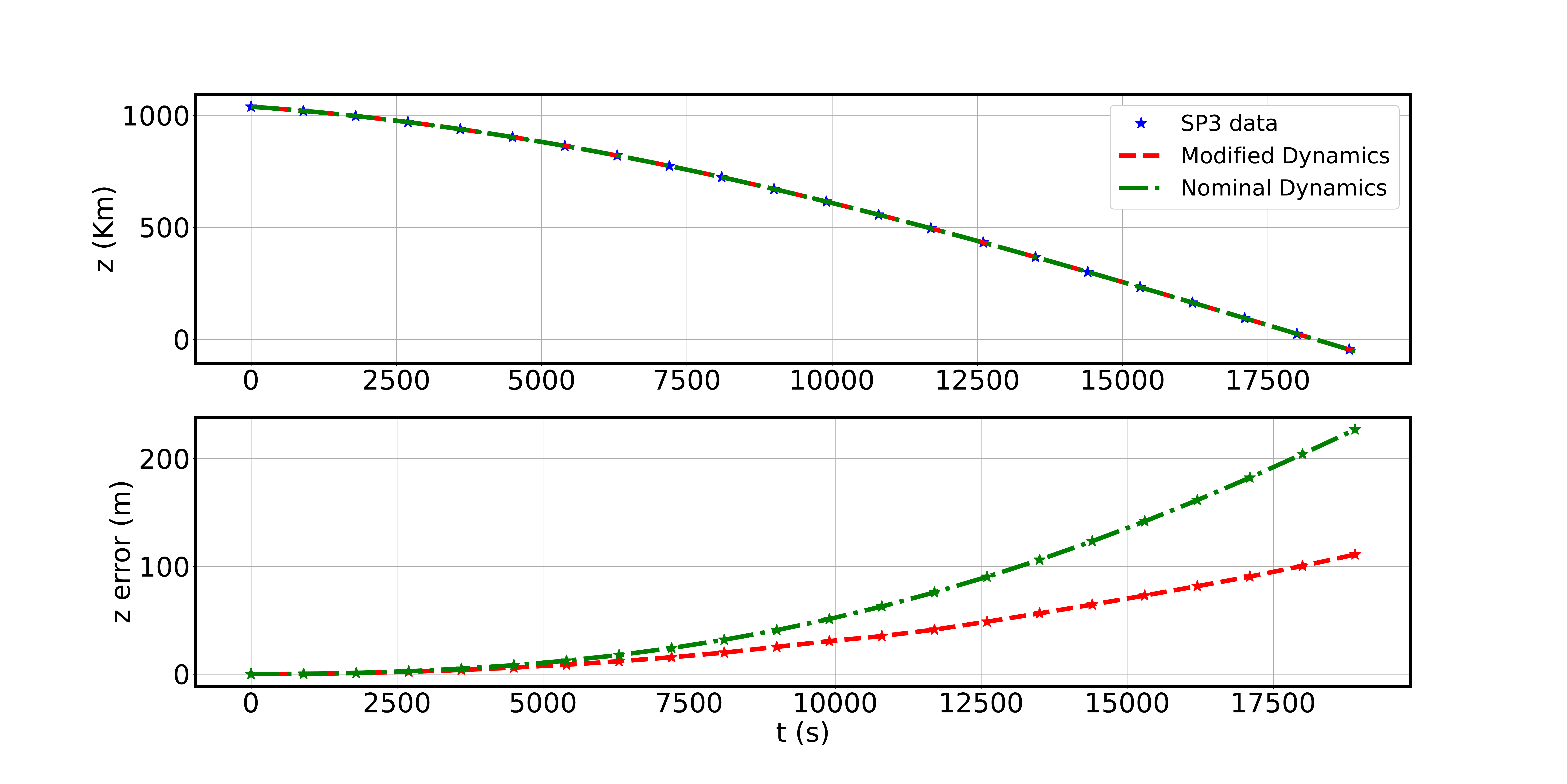}
 \caption{\label{figure:Figure1} Predicted satellite coordinates in ICRF using the proposed approach and by propagating the nominal dynamics. Predicted values are compared against the SP3 data and the absolute value of difference is plotted as coordinate error. Error plots are interpolated between the marked points where actual differences are computed.}
\end{figure}

In the trapezoidal scheme, the Lagrange multiplier vector at the $(k+1)^{th}$ time step $\boldsymbol \lambda^{k+1}$ has to be selected from the $\lambda$-dataset on the basis of the position $\boldsymbol r^{k+1}$. At the $k^{th}$ time step, $\boldsymbol r^{k+1}$ is not known and, as a matter of fact, it is the very quanity we are predicting. We use forward Euler method to compute $\boldsymbol r^{k+1}$ which in turn is used to determine $\boldsymbol \lambda^{k+1}$ from the $\lambda$-dataset. This $\boldsymbol \lambda^{k+1}$ is then used in equation \ref{TM} to compute $\boldsymbol{v}^{k+1}$.\par
\begin{align}
v_i^{k+1}=&v_i^k+\frac{h}{2}\Bigg[\frac{-GMx_i^k}{{[(x_1^k)}^2+{(x_2^k)}^2+{(x_3^k)}^2]^{3/2}}+\lambda_i^k\Bigg] \nonumber \\ &+\frac{h}{2}\Bigg[\frac{-GMx_i^{k+1}}{{[(x_1^{k+1})}^2+{(x_2^{k+1})}^2+{(x_3^{k+1})}^2]^{3/2}}+\lambda_i^{k+1}\Bigg] \label{TM}
\end{align}
For notational convenience while deriving the equations,  the position coordinates were denoted as $x_1$, $x_2$ and $x_3$. To present the results of the numerical example that follows, we adapt to the IGS SP3 notation and denote $x_1$, $x_2$ and $x_3$ by x, y and z respectively.
In the following numerical example, orbit of a geostationary satellite of BeiDou Navigation Satellite System (BDS) satellite constellation is predicted. Precise ephemeris for BDS satellite C05 obtained for a period starting from 10 Dec 2015 upto 19 Dec 2015 (GPS time) from the IGS final orbit product distributed in SP3 format is used as the historical dataset. This dataset is used to compute the values of $\boldsymbol{\lambda}$ by solving the DAE system \ref{8} and \ref{11} as described earlier in the text using numerical scheme \ref{sc1}-\ref{16} to create the $\lambda$-dataset. Satellite positions are predicted by integrating the modified evolution model \ref{11} for a period of \SI{19000}{s} ($\approx$ 5.27 hours) starting at 0 GPS hours next day (Dec 20 2015) to give predicted coordinates with \SI{1}{s} time gap. Absolute value of the difference between the predicted coordinates and the SP3 precise coordinates transformed to ICRF gives the absolute coordinate errors. These errors are computed at \SI{15}{min} intervals, which corresponds to the spacing between the epochs of SP3 precise ephemeris. Figure \ref{figure:Figure1} shows a comparison of the predicted satellite coordinates with precise ephemeris and the corresponding errors. At the end of 2 hours, the predicted coordinates have an absolute error of \SI{34.448}{m}, \SI{51.319}{m} and \SI{15.649}{m} in x, y and z directions respectively. The errors in x, y and z coordinates increase to \SI{56.841}{m}, \SI{532.966}{m} and \SI{110.981}{m} when the duration of prediction is \SI{18900}{s}($\approx$ \SI{5.25}{hours}).\par 
Prediction error is also compared with corresponding errors when direct integration of equation \ref{12} is performed using the Verlet scheme~\cite{Hairer}. A smaller time step of \SI{0.1}{s} is used in this case. The absolute prediction error in the x, y and z directions for the nominal gravitational model (i.e. equation \ref{12}) is  \SI{454.570}{m},  \SI{289.344}{m} and \SI{24.212}{m} respectively with respect to the precise ephemeris at the end of 2 hours. When integrated for \SI{18900}{s}, these absolute errors are \SI{3728.944}{m}, \SI{406.571}{m} and \SI{227.027}{m} respectively.\par
\begin{figure}
\centering
	\includegraphics[width=1\linewidth]{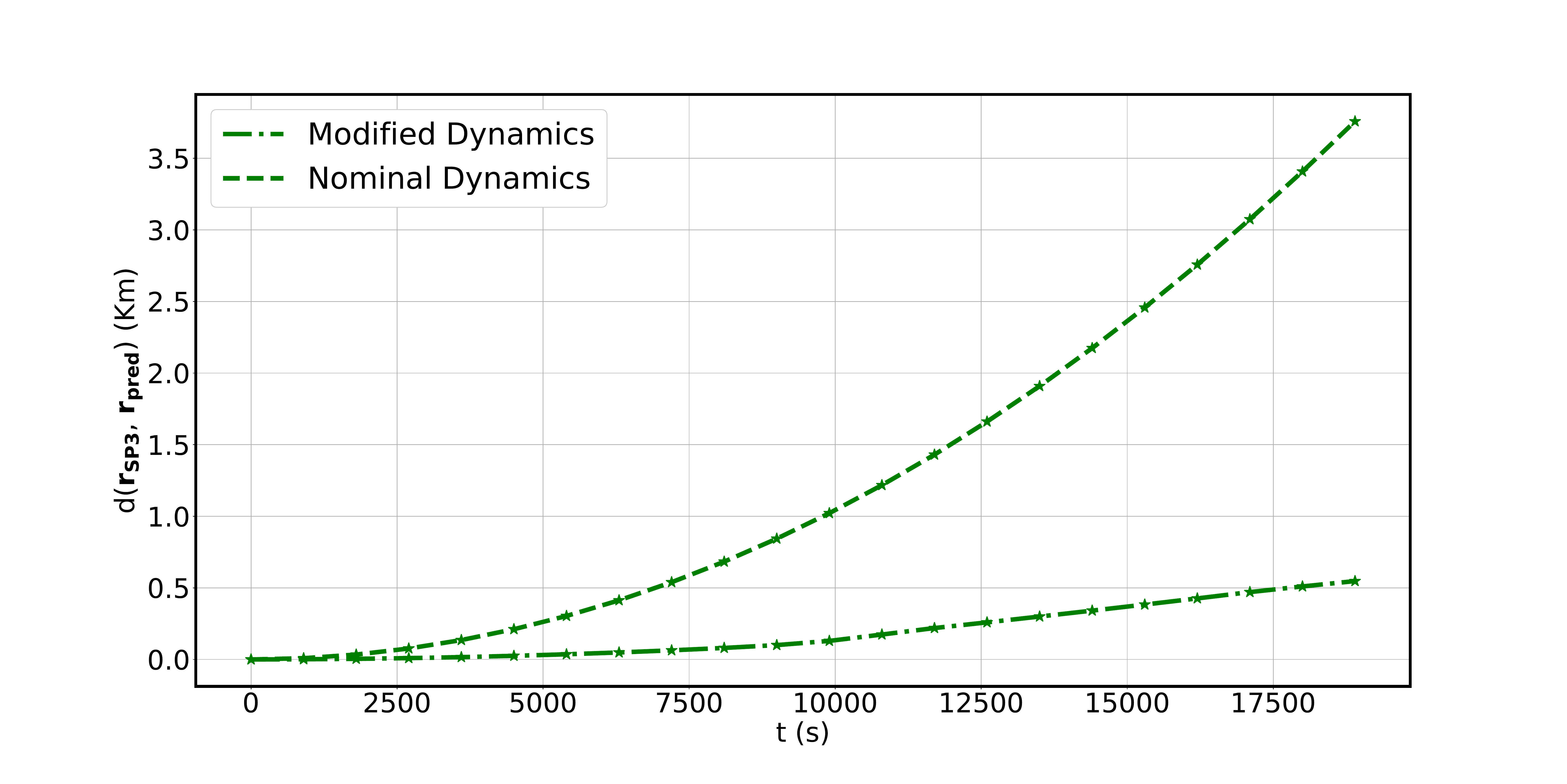}
	\caption{Comparison of Euclidean distance, $d(\boldsymbol{r_{SP3}},\boldsymbol{r_{pred}})$ between the precise (SP3) coordinate vector($\boldsymbol{r_{SP3}}$) and the predicted coordinate vector($\boldsymbol{r_{pred}}$) using the modified and the nominal dynamics model. Plots are interpolated between the marked points where precise coordinates are available for comparison.} \label{fig1.1}
\end{figure}

Figure \ref{fig1.1} shows the time variation of the Euclidean distance $d(\boldsymbol{r_{SP3}}, \boldsymbol{r_{pred}})$ between the predicted position vector $\boldsymbol{r_{pred}}$ and the SP3 precise position vector $\boldsymbol{r_{SP3}}$ of the satellite. At the end of 2 hours, by using the dynamics model modified with Lagrange multiplier term, $d(\boldsymbol{r_{SP3}}, \boldsymbol{r_{pred}})$ is \SI{63.759}{m}. For the same interval, when nominal gravitional dynamics equation is integrated, the Euclidean distance between the predicted and precise position of satellite is \SI{539.389}{m}. For \SI{5.25}{h} prediction period, Euclidean distance for the modified model is \SI{547.357}{m} while for the nominal model it is \SI{3757.907}{m}. This Euclidean distance is the straight line distance between the precise and the predicted position vectors of the satellite and thus gives a measure of error in the predicted position of the satellite. We observe that in comparison with the satellite position computed using the nominal gravitational model, the predicted position using the modified model is much nearer to its precise ephemeris position.
\\
\section{Temperature prediction in one-dimensional heat conduction}
Data for the satellite orbit prediction example of the last section has a characteristic of repeating orbits. More than one $\lambda$ evaluations can be performed 
for the same sector of space because of the cyclic nature of satellite orbits. In other words, the satellite returns to the approximate position for which $\lambda$ has been evaluated using the historical data. 
The temperature during heat conduction approaches a steady-state and unlike the satellite orbits a similar value of temperature is not reached at the same point on the object again in one experiment. Therefore, a nearest neighbour approach cannot be used for selecting $\lambda$ during prediction in this case. Using the one-dimensional heat conduction example in this section, we illustrate that the same formulation can be adapted to different problems with suitable problem-specific improvisations.

\subsection{Heat equation with constraint}
For one-dimensional heat conduction through a metal rod, temperature u is a function of both the distance from the end point x
and time t. If the dynamics in this case is subjected to k constraints of type $\boldsymbol{M}(u)=0$ with k mirror-image system constraints $\boldsymbol{M^{'}}(u^{*})=0$, then the Lagrangian $L^{'}$ is given by:
\begin{equation}
L^{'}=-\frac{\partial u}{\partial x}\frac{\partial u^*}{\partial x} -\frac{1}{2}a^2\Big(u^*\frac{\partial u}{\partial t}-u\frac{\partial u^*}{\partial t}\Big)+\sum_{l=1}^{k}\lambda_l(x,t)M_l(u)+\sum_{l=k+1}^{2k}\lambda_l(x,t)M_l^{'}(u^{*}) \label{21}.
\end{equation}
Lagrangian for heat equation without constraints and mirror-image system approach for dissipative systems can be found in \parencite[][chapter 3]{Morse}.
Using equation \ref{2} with $h_1=u$, $h_2=u^{*}$, $\omega_1=x$ and $\omega_2=t$, we get,
\begin{equation}
 a^2\frac{\partial u}{\partial t}=\frac{\partial^2 u}{\partial x^2}+ \sum_{l=k+1}^{2k} \lambda_l(x,t) \frac{\partial}{\partial u^*}(M_l^{'}(u^*)) \label{22}.
\end{equation}
If there is single constraint of type $u(x,t)-p(x,t)=0$ where $p(x,t)$ is some known function, then corresponding constraint in 
the mirror-image system will also be of the form $u^{*}(x,t)-p^{'}(x,t)=0$ and \ref{22} can be expressed as:
\begin{equation}
 a^2\frac{\partial u}{\partial t}=\frac{\partial^2 u}{\partial x^2}+\bar{\lambda}(x,t) \label{23},
\end{equation}
where $\bar{\lambda}$ is the Lagrange multiplier corresponding to the single constraint.
\subsection{Heat conduction numerical example}
In section 4.1, we derived the equation for constrained heat-conduction dynamics. We now use data from a heat conduction experiment to evaluate the
Lagrange undetermined multipliers in order to estimate the missing dynamics in the nominal model of heat conduction for this experiment.
The data set used in this example is obtained from \cite{AndrewsAdv}. The setup described by the experimenters is as follows: A radially insulated aluminium rod is dipped in ice bath at one end while the other end is kept at the room temperature. Temperature is measured at 10 different locations on the rod with \SI{2}{s} interval between measurements. Specifications of the rod are as follows: 
  Length : \SI{0.306}{m};
  Thermal conductivity (theoretical), $k$ : \SI{209}{W/mK}; 
  Density (theoretical), $\rho$ : \SI{2763.14}{Kg/m^3};
  Specific heat capacity (theoretical), $c_p$ : \SI{900}{J/KgK};
  Temperature at ends : \SI{273.15}{K} and \SI{292.65}{K}.

Let the x-axis of the coordinate system be placed along the length of the rod and let the rod end dipped in ice bath be the origin $(x=0)$. The nominal model for this experiment is the usual heat equation without constraints, given as:
  \begin{equation}
		  \frac{\partial u}{\partial t}=\alpha \frac{\partial ^2 u}{\partial x^2} \label{24}.
  \end{equation}
 Here, $u(x,t)$ is the temperature of rod at a location $x$ and time $t$, $\alpha$ is the thermal diffusivity given by $\frac{k}{c_p\rho}$. 
Let $x_0$ denote the origin and let $x_1,x_2,\dots,x_{n-1}$ be the points on the x-axis placed along the rod, where the temperature measurements are made.
Let $x_n$ be the other end point maintained at the room temperature of \SI{292.65}{K}. If $u_0$ and $u_n$ represent the temperatures at the end points of the rod, then the boundary conditions for this experiment is given by
\begin{align}
u_0&=\SI{273.15}{K} \nonumber \\
u_n&=\SI{292.65}{K} \label{BC}
\end{align}
\begin{figure}
	\centering
	\includegraphics[width=\textwidth]{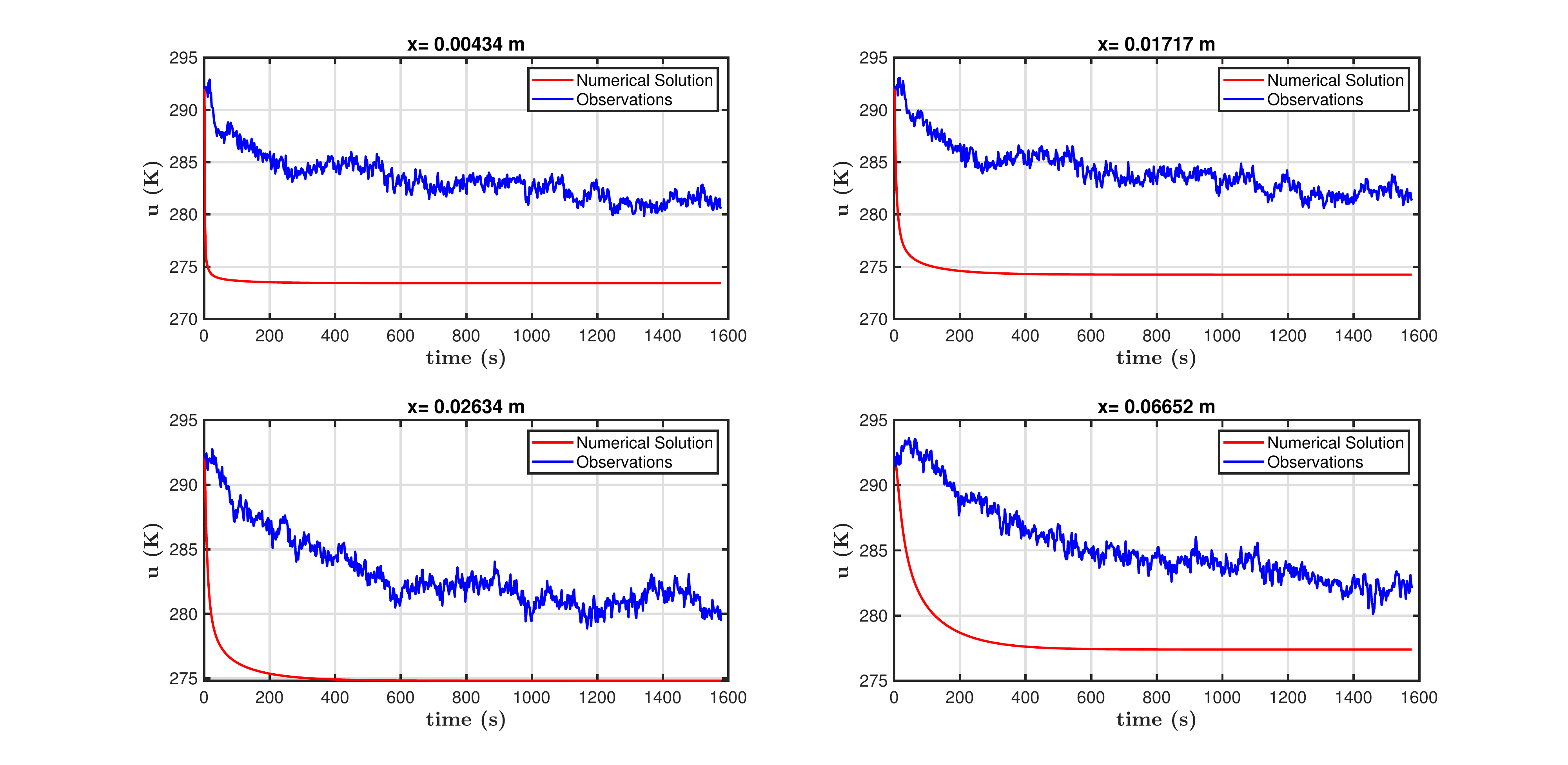} 
	\includegraphics[width=\textwidth]{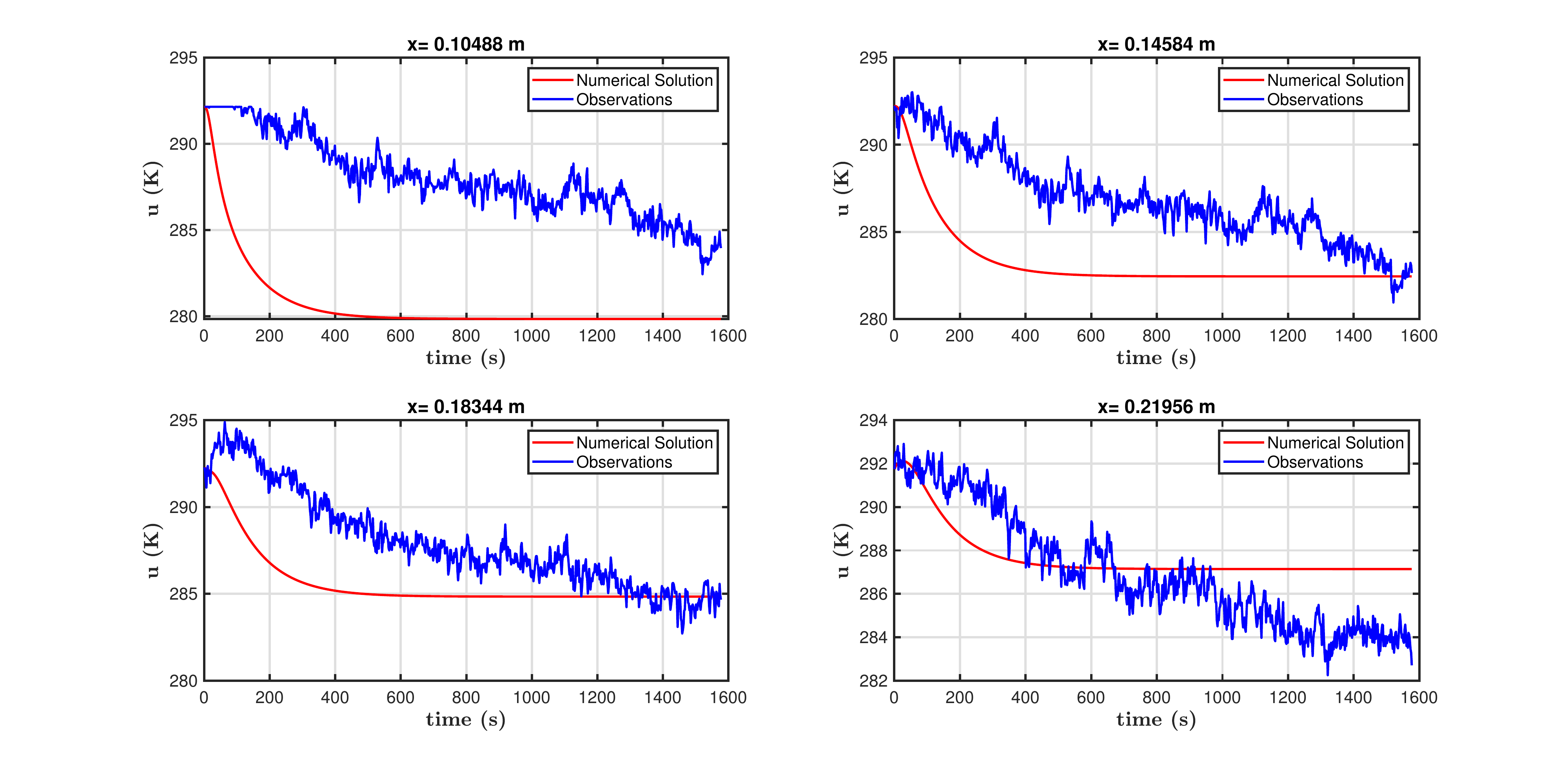}
	\includegraphics[trim={0 8.3cm 0 0},clip,width=\textwidth,height=3.9cm,angle=0]{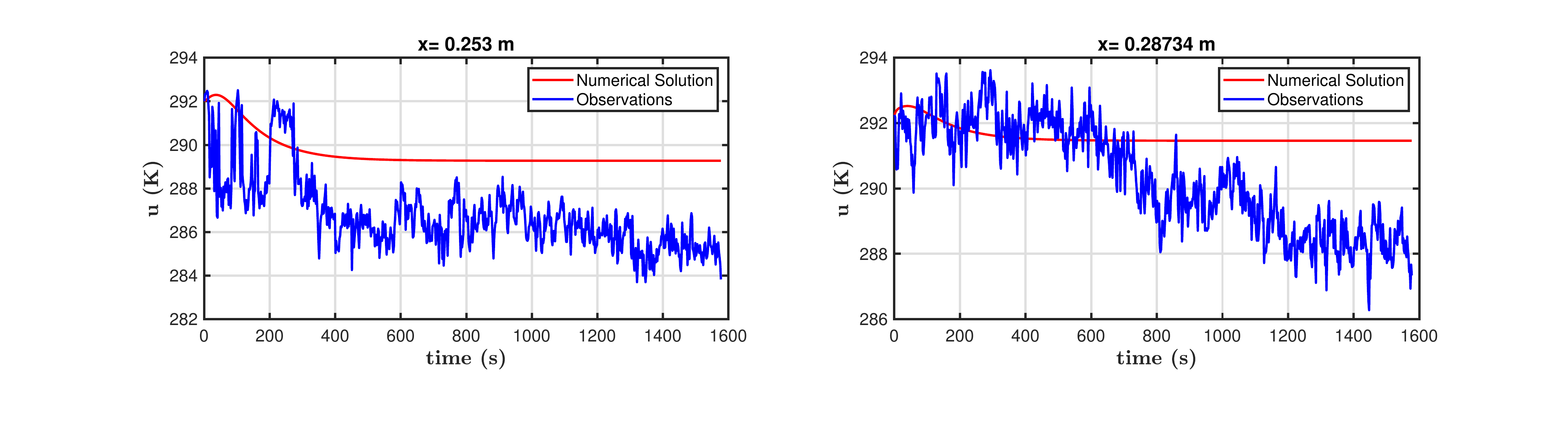}
	\caption{Comparison of numerical solution of the heat equation obtained using the measured thermal diffusivity (red) with the experimental data (blue) for all measurement nodes. x coordinate value at the top of each plot determines the location of the measurement node.}\label{fig1a}
\end{figure}
Nominal heat equation \ref{24} is solved numerically with boundary conditions \ref{BC} and initial condition determined by the first temperature measurement at all measurement locations. Figure \ref{fig1a} compares the solution of the heat equation \ref{24} with the observed values of temperature at the 10 measurement nodes. It can be observed that the numerical solution quickly reaches a steady state as opposed to the experimental data in the measurement duration. Also, the
numerical solutions are smooth while the observations are noisy.\par
Similar to the orbit prediction problem, equation \ref{23} along with the measurement data constraints is used to determine the 
Lagrange multipliers at each time-step and measurement locations on the rod.\\
Dividing \ref{23} by $a^2$ and denoting $\frac{1}{a^2}$ by $\alpha$ and $\frac{\bar\lambda}{a^2}$ by $\lambda$ gives,
\begin{equation}
 \frac{\partial u}{\partial t}=\alpha \frac{\partial ^2 u}{\partial x^2}+\lambda \label{25}.
\end{equation}
Equation \ref{25} has a form similar to \ref{24} with $\lambda$ as an additional term.\par
In order to solve for $\lambda$, equation \ref{25} along with the constraints is discretized in space coordinates to obtain a DAE system. Let $u_i$, $\lambda_i$ be the
temperature and the value of Lagrange Multiplier at position $x_i$ and let $h_{i1}=x_{i+1}-x_i$ (for $i<n$) and 
$h_{i2}=x_{i}-x_{i-1}$ (for $i>0$), $i \in \{0,1,\dots, n\}$. Discretizing \ref{25} in space coordinates using central difference with unequal discretization intervals, we get:
 \begin{equation}
		 \frac{du_i}{dt}=2\alpha\frac{h_{i1}u_{i-1}-(h_{i1}+h_{i2})u_{i}+h_{i2}u_{i+1}}{h_{i1}h_{i2}(h_{i1}+h_{i2})}+\lambda_{i}, \label{25a}
\end{equation} 
for $i \in \{1,2,\dots, n-1\}$ . Equation \ref{25a} represents a set of ODEs with time t as the indepedent variable.\par
In vector notation, for all i's \ref{25a} together with the boudary conditions \ref{BC} can be written as,
\begin{equation}
 \boldsymbol {\dot{u}=L u+D^T\lambda}, \label{27a}
\end{equation}
where ,
\begin{multicols}{3}
\begin{equation*}
 \boldsymbol{u}={\begin{bmatrix}
	      u_0 & u_1 & \dots & u_n
             \end{bmatrix}}^{'},
 \end{equation*}\break
\begin{equation*}
 \boldsymbol{\dot{u}}=\frac{d\boldsymbol{u}}{dt},
\end{equation*}\break
\begin{equation*}
 \boldsymbol{\lambda}={\begin{bmatrix}
                        0 & \lambda_1 & \lambda_2 & \dots & \lambda_{n-1} & 0
                       \end{bmatrix}}^{'},
\end{equation*}
\end{multicols}

\begin{equation} L= \newline
\scalemath{1}
{2\alpha
\begin{bmatrix}
         0 & 0 & \dots & \dots & \dots & 0\\
         a_1 & b_1 & c_1 & 0 & \dots & 0 \\
	 0 & a_2 & b_2 & c_2 & \dots & 0 \\
	 \vdots & \vdots & \vdots & \vdots & \vdots & \vdots \\
	 \vdots & \vdots & \vdots & \vdots & \vdots & \vdots\\
	 0 & 0 & \dots & a_{n-1} & b_{n-1} & c_{n-1} \\
	 0 & 0 & \dots & 0 & 0 & 0
	 
   \end{bmatrix},
   }
\end{equation}
with
\begin{align}
a_i=& \frac{1}{h_{i2}(h_{i1}+h_{i2})}, \nonumber\\
b_i=&-\frac{1}{h_{i1}h_{i2}},\nonumber\\
c_i=&\frac{1}{h_{i1}(h_{i1}+h_{i2})}\nonumber;
\end{align}
and 
\begin{align} 
\boldsymbol{D}=\boldsymbol{I_{(n+1)\times (n+1)}}. \label{30}
\end{align}
Temperature measurements are available at discrete locations, and the constraint equations $u(x,t)-p(x,t)=0$ take the form $u(x_i,t)-p(x_i,t)=0$ with $i\in\{1,2,\dots,n-1\}$. \par
 The observation vector $\boldsymbol{Y}(t)=(y_0,y_1,\dots,y_n)^T$ forms the following system of algebraic equations:
\begin{equation}
  \boldsymbol{Du}-\boldsymbol{Y}(t)=\boldsymbol 0. \label{30}
\end{equation}

It should be noted that the matrix $\boldsymbol{D}$ is identity in this case owing to direct observation of temperatures, but in a more general form where the observables are some function of the temperature, $\boldsymbol{D}$ will be the Jacobian matrix  as evident from \ref{22} . To be mindful of this, we denote it by $\boldsymbol{D}$ instead of using the usual identity matrix notation $\boldsymbol{I}$.
 Equations \ref{27a} and \ref{30} form a system of Hessenberg Index-2 Differential Algrebraic Equations \cite{Petzold}. Equation \ref{30} takes care of the boundary conditions \ref{BC} by having $y_0=\SI{273.15}{K}$ and $y_n=\SI{292.65}{K}$ for all t. \newline 
For temporal discretization, let $t_0=\SI{0}{s}$, $\Delta t$ be the time step and $t_k=t_0+k\Delta t$ with $k$ assuming integer values. Let $u_{i}^k$ and $\lambda_{i}^k$ be the temperature and the value of Lagrange multiplier at the position $x_i$ and time $t_k$.
By Backward Euler Method \ref{25a} can be discretized as:
\begin{align}\label{30.1}
 \frac{u_i^k-u_i^{k-1}}{\Delta t}=2\alpha\frac{h_{i1}u_{i-1}^k-(h_{i1}+h_{i2})u_i^k+h_{i2}u_{i+1}^k}{h_{i1}h_{i2}(h_{i1}+h_{i2})}+\lambda_{i}^k
\end{align}

\begin{align}\label{31}
\implies u_i^k-u_i^{k-1}=2\alpha \Delta t\frac{h_{i1}u_{i-1}^k-(h_{i1}+h_{i2})u_i^k+h_{i2}u_{i+1}^k}{h_{i1}h_{i2}(h_{i1}+h_{i2})}+\Delta t \lambda_i^k \nonumber \\
\implies
 -2\alpha \Delta t\frac{h_{i1}u_{i-1}^k}{h_{i1}h_{i2}(h_{i1}+h_{i2})}+\Big(1+2\alpha \Delta t\frac{(h_{i1}+h_{i2})}{h_{i1}h_{i2}(h_{i1}+h_{i2})}\Big)u_i^k \nonumber \\
 -2\alpha \Delta t\frac{h_{i2}u_{i+1}^k}{h_{i1}h_{i2}(h_{i1}+h_{i2})}=u_i^{k-1}+\Delta t \lambda_i^k. 
\end{align}
Let the matrix $\boldsymbol{\tilde{L}}$ be of the following form,
\begin{equation*}
{
\begin{bmatrix}
         1 & 0 & \dots & \dots & \dots & 0\\
         a_1 & b_1 & c_1 & 0 & \dots & 0 \\
	 0 &  a_2 & b_2 & c_2  & \dots & 0 \\
	 \vdots & \vdots & \vdots & \vdots & \vdots & \vdots \\
	 \vdots & \vdots & \vdots & \vdots & \vdots & \vdots\\
	 0 & 0 & \dots &  a_{n-1} & b_{n-1} & c_{n-1} \\
	 0 & 0 & \dots & 0 & 0 & 1
	 
   \end{bmatrix}
   }
\end{equation*}
with
\begin{align*}
a_i&=  -2\alpha \Delta t \frac{1}{h_{i2}(h_{i1}+h_{i2})} \\
b_i&=  1+2\alpha \Delta t\frac{1}{h_{i1}h_{i2}} \\
c_i&=  -2 \alpha \Delta t\frac{1}{h_{i1}(h_{i1}+h_{i2})}.
\end{align*}
Combining \ref{31} for all i and the boundary conditions in a single vector equation we get, 
		 \begin{equation}
                  \boldsymbol{\tilde{L}u^k}=\boldsymbol{u}^{k-1}+\Delta t \boldsymbol{D^{T}\lambda^k}, \label{32}
                 \end{equation}
where $\boldsymbol{u^k}$ and $\boldsymbol{\lambda^k}$ are $\boldsymbol{u}$ and $\boldsymbol{\lambda}$ respectively evaluated at time $t_k$. Constraint equation \ref{30} evaluated at $t_k$ is combined with \ref{32} to give,
\begin{equation}\label{33a}
 \begin{bmatrix}
 
 \boldsymbol{\tilde{L}} & -\Delta t \boldsymbol{D^T} \\
  \boldsymbol{D} & \boldsymbol{0}
 \end{bmatrix}
\begin{bmatrix}
 \boldsymbol{u}^k \\
 \boldsymbol{\lambda}^k
\end{bmatrix}
=
\begin{bmatrix}
 \boldsymbol{u}^{k-1}\\
 \boldsymbol{Y}(t_k)
\end{bmatrix}.
\end{equation}
The measurement data is partitioned into two sets:
\begin{enumerate}
\item The first partition comprises of the first 600 observations till the measurement time $\SI{1198.9}{s}$. This partition is used to compute $\lambda_i^t$, which is used to estimate the additional term in the evolution model of the system to match the observations. 
\item The second partition consists of the rest 189 observations starting at measurement time $\SI{1200.9}{s}$ till the time $\SI{1576.9}{s}$. The quality of prediction obtained using the modified evolution model is assessed by comparing it with the observations in this partition.
\end{enumerate}
Observations are available at every 2 seconds in both the partitions (except the first observation in first partition which is at time \SI{0.9}{s} ). Equation \ref{33a} is solved with $\Delta t=\SI{2}{s}$ and $\boldsymbol{Y}(t_k)$ taking values from the first data partition so that an observation is present at each time step of computation. 
\begin{figure}
	\includegraphics[width=\textwidth]{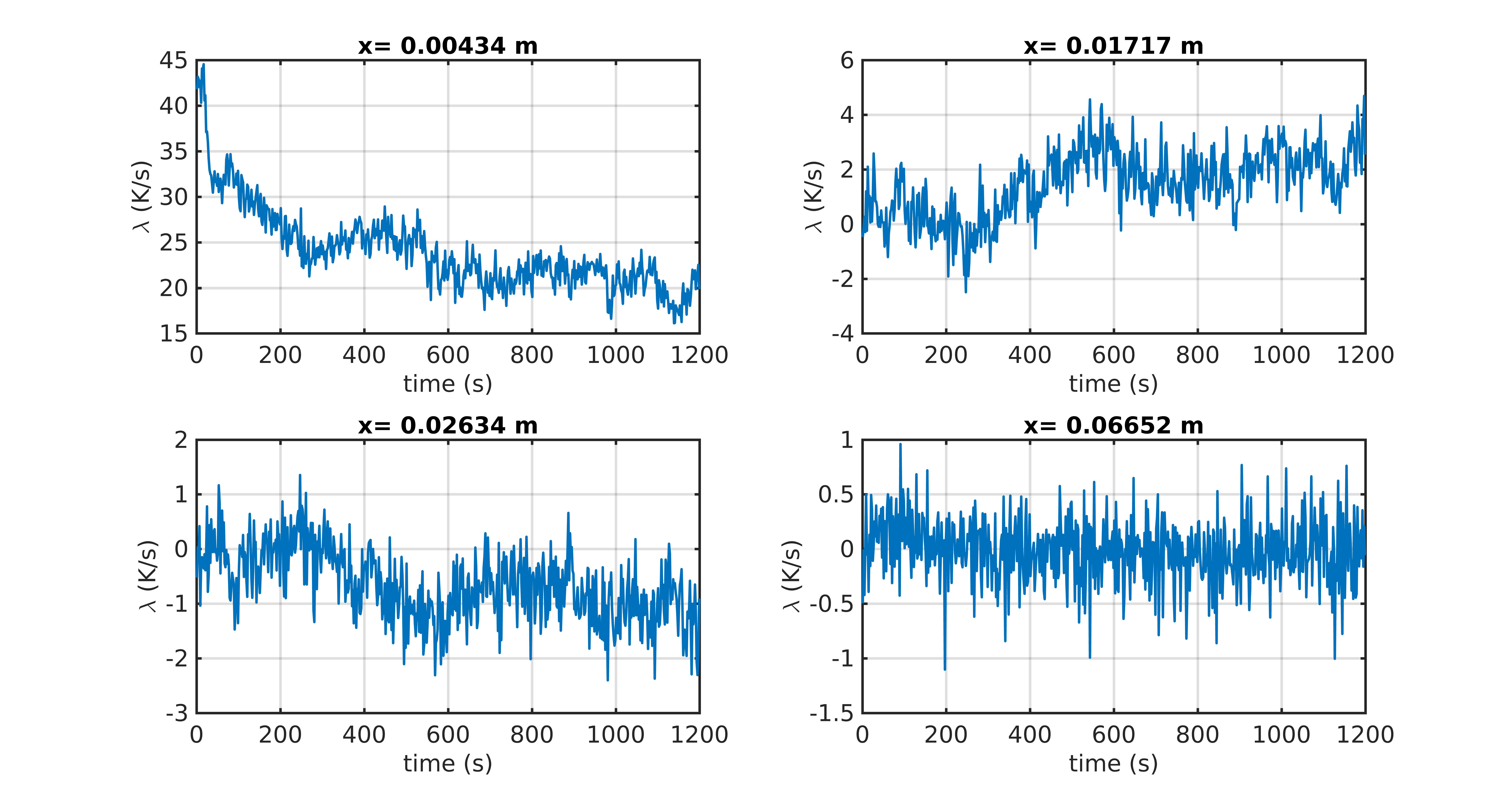}
	\includegraphics[width=\textwidth]{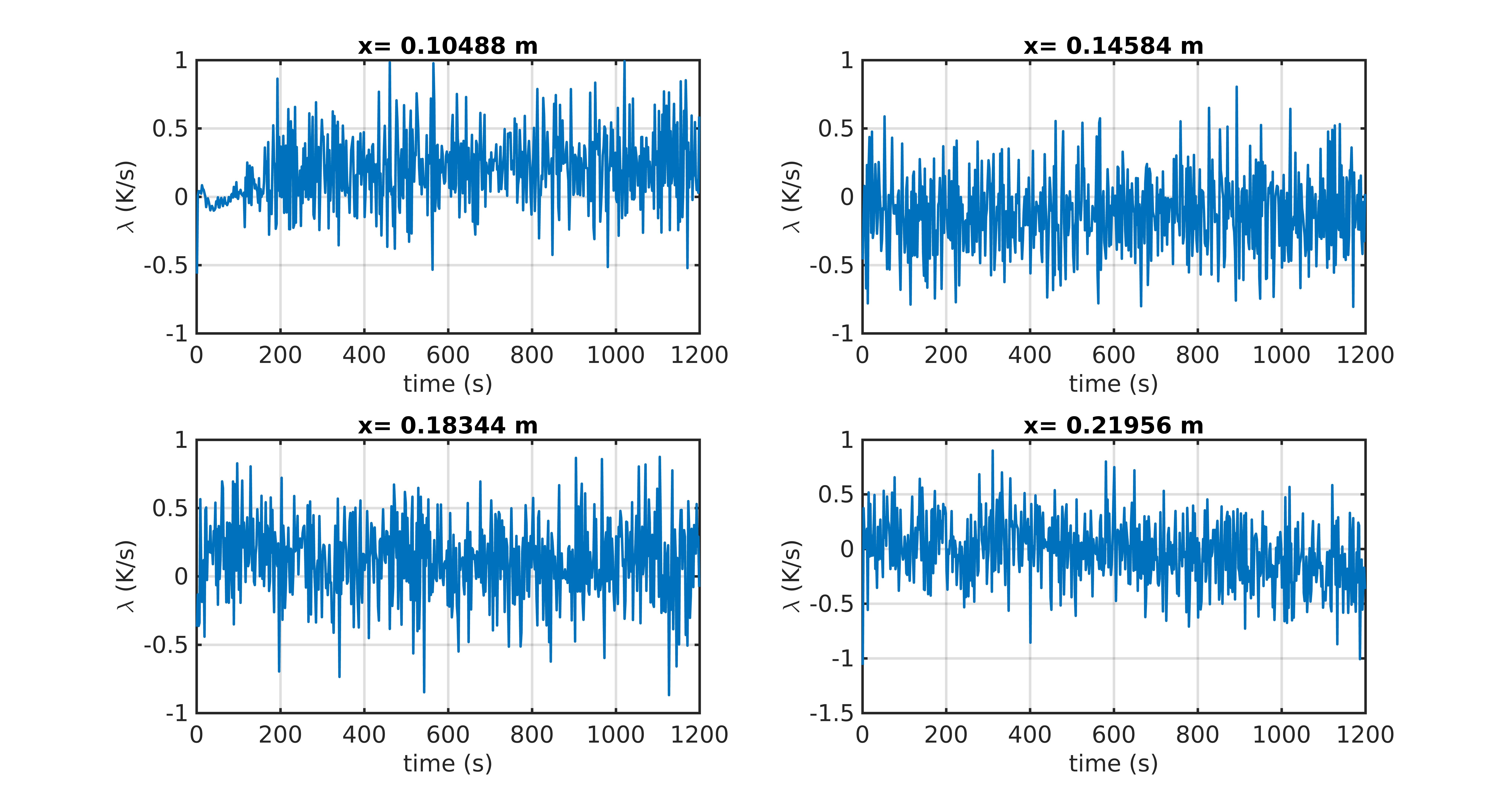}
	\includegraphics[trim={0 70cm 0 0},clip,width=\textwidth,height=3.8cm,angle=0]{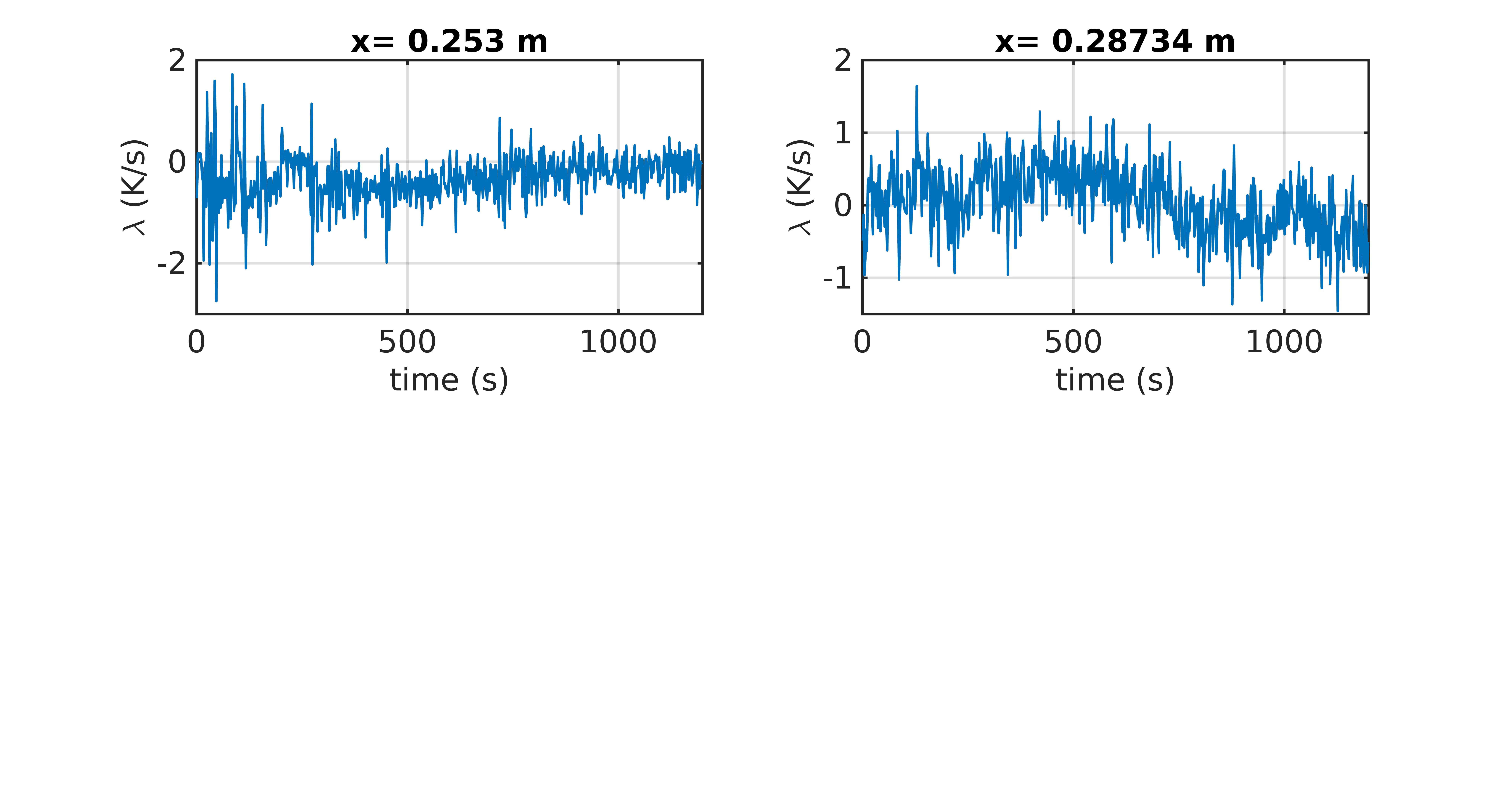}
	\caption{Lagrange multiplier $\lambda$ plotted against time (t) for all measurement nodes.  x coordinate value at the top of each plot gives the location of the measurement node. For instance, x=0.00434 is the location $x_1$, the first measurement node and the corresponding $\lambda$ is $\lambda_1$. } \label{fig5}
\end{figure}
\begin{figure}
	\includegraphics[width=\textwidth]{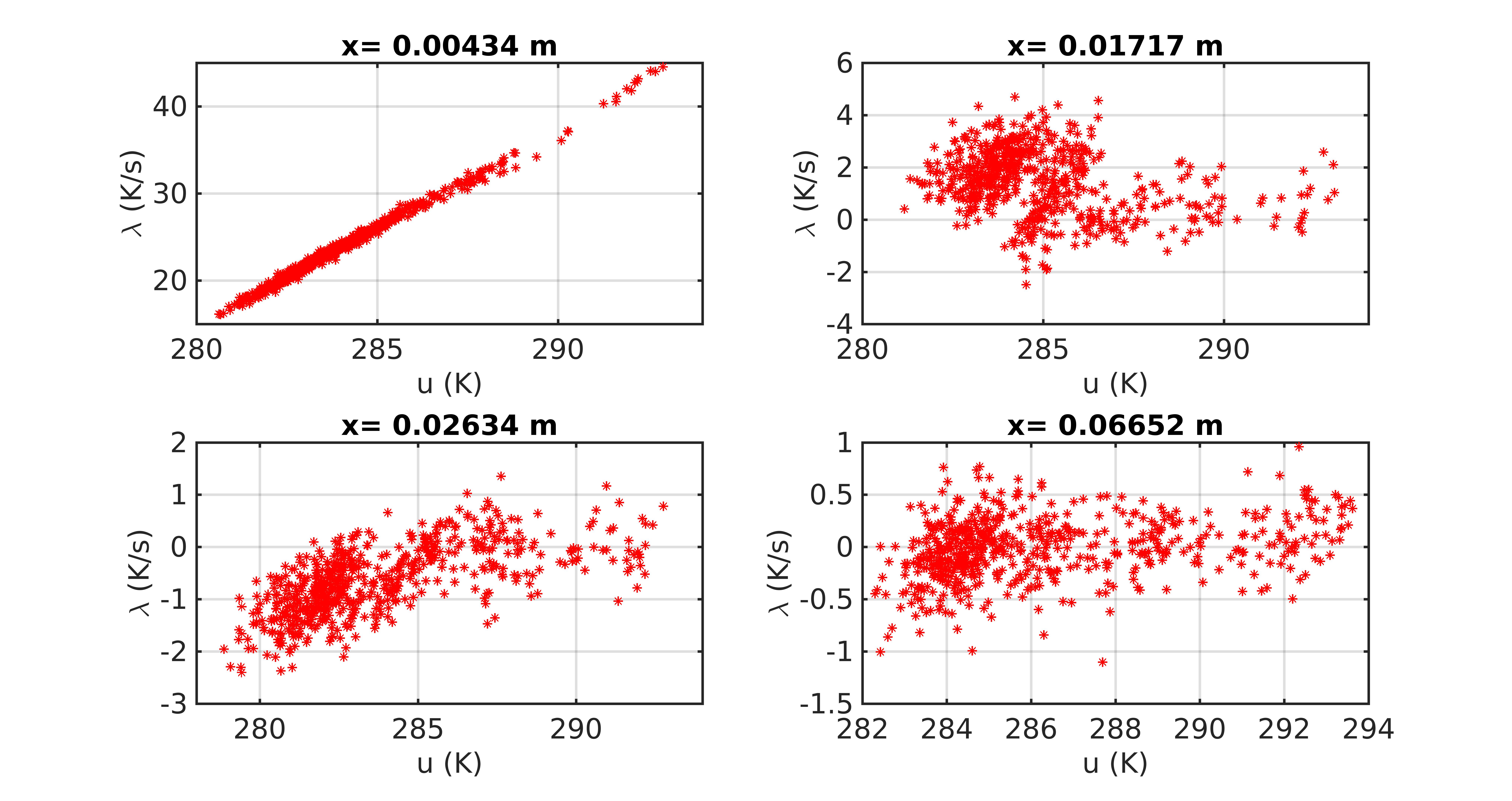}
	\includegraphics[width=\textwidth]{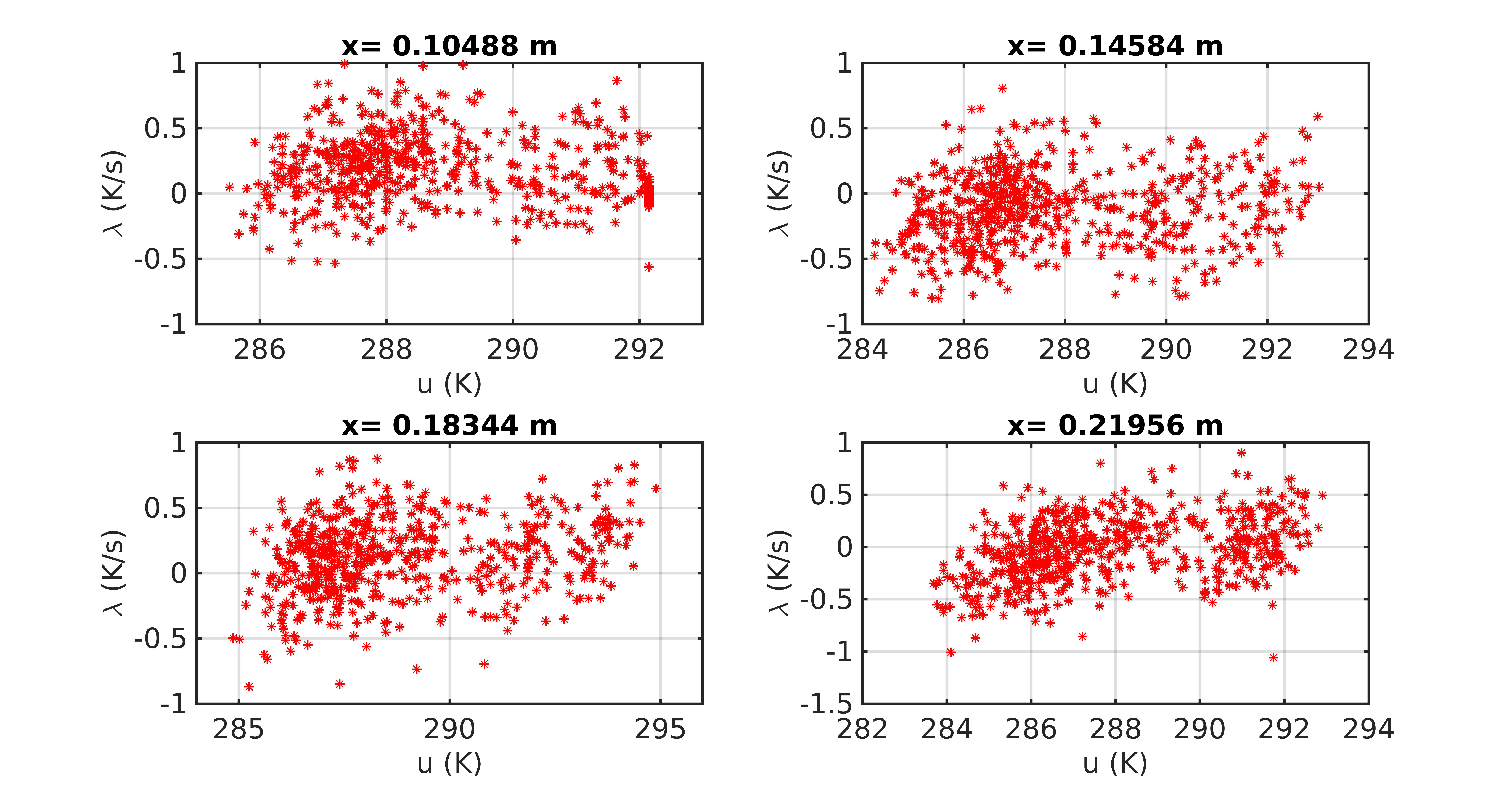}
	\includegraphics[trim={0 70cm 0 0},clip,width=\textwidth,height=3.8cm,angle=0]{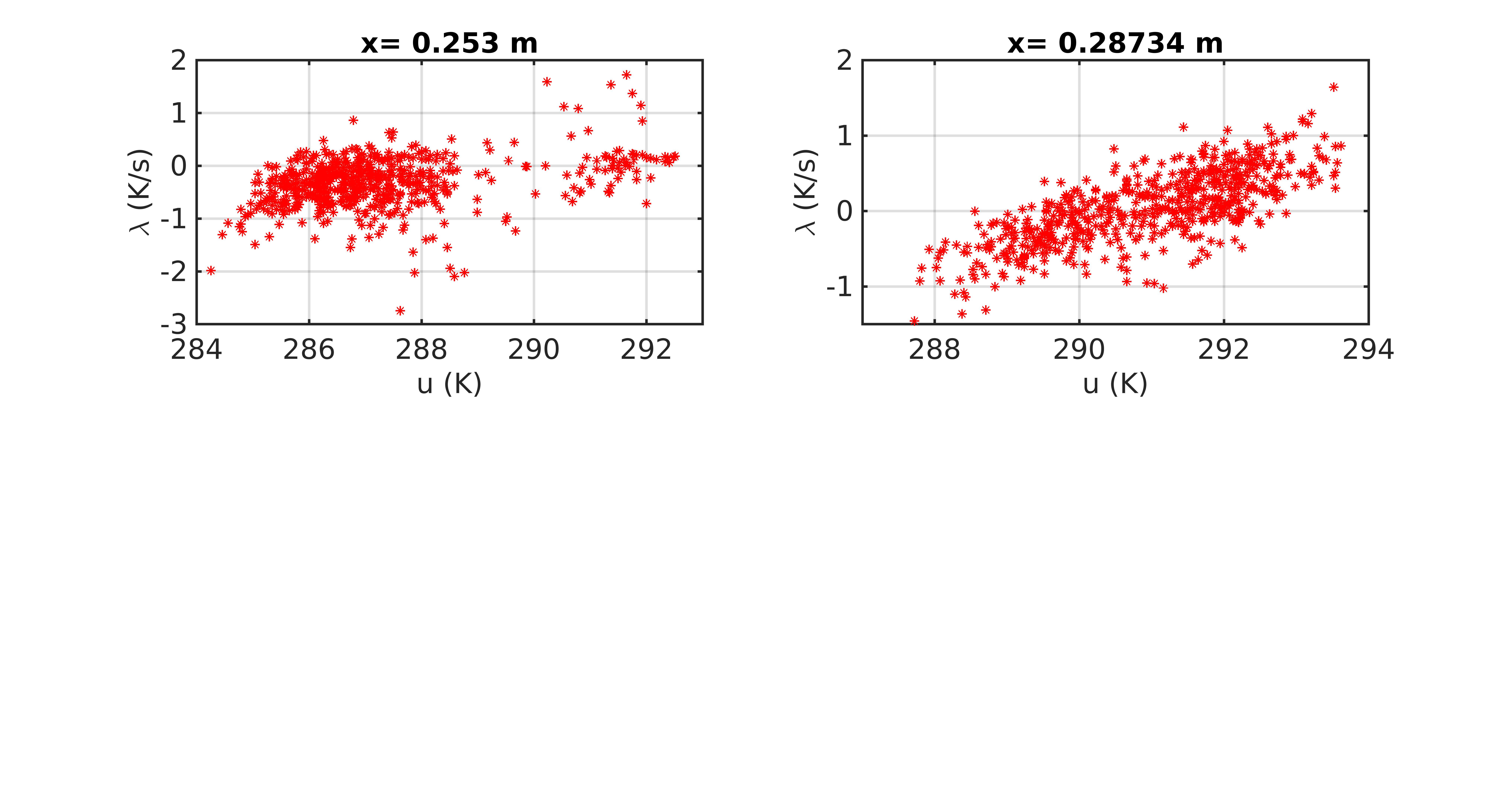}
	\caption{DAE system is solved for Lagrange multiplier $\lambda$ for all t at each measurement node. $\lambda$ variation against measured temperature u, at each node is presented here. As before, the x coordinate value at the top of each plot gives the location of the measurement node.} \label{fig6}
\end{figure}

Figures \ref{fig5} and \ref{fig6} show the plots of $\lambda_i$ with time and the observed temperature $u_i$ for $i\in\{1,2,\dots,n-1\}$. There is no obvious pattern of variation of $\lambda_i$'s in both the figures. However, in figure \ref{fig6}, it is observed that $\lambda$ lies roughly on a straight line with temperature for nodes at x=0.00434 mm and x=0.28734 mm. These are the measurement nodes $x_1$ and $x_{n-1}$, which are adjacent to the boundary of the rod. \par
Expressing the partial derivative of u with respect to x using backward difference approximation and letting $\Delta x=x_{i+1}-x_i$, we get,
\begin{equation}\label{33}
	   \frac{\partial u_{i+1}}{\partial x} \approx \frac{u_{i+1}-u_i}{\Delta x}=\frac{u_{i+1}}{\Delta x}-\frac{u_i}{\Delta x}.
\end{equation}

      If in \ref{33} $\frac{u_{i}}{\Delta x}=c_1$, where $c_1$ is some constant, then \ref{33} can be written as,
      \begin{equation*}
 	      \frac{\partial u_{i+1}}{\partial x} \approx \frac{u_{i+1}-u_i}{\Delta x}=\frac{u_{i+1}}{\Delta x}-c_1.
      \end{equation*}

     Additionally, if $\frac{\partial u_{i+1}}{\partial x}=\gamma \lambda_{i+1}+c_2$, where $\gamma$ and $c_2$ are constants, then $\Delta x\gamma \lambda_{i+1}+\Delta x(c_1+c_2) \approx u_{i+1}$. Since, $\Delta x$,$\gamma$, $c_1$ and $c_2$ are all constants, there is a linear relationship between $\lambda_{i+1}$ and $u_{i+1}$. This suggests that if there is a linear relationship between $\frac{\partial u}{\partial x}$ and $\lambda$ at all nodes, then there will also be a linear relationship between $u_1$ and $\lambda_1$ i.e. $u$ and $\lambda$ at the first node.\par
Similar arguments can be made for the second partial derivative of u with respect to x as follows. Using the central difference formula as in \ref{25a}, we have,
      \begin{equation}\label{36}
       \frac{\partial^2 u_i}{\partial x^2} \approx 2\frac{h_{i1}u_{i-1}-(h_{i1}+h_{i2})u_{i}+h_{i2}u_{i+1}}{h_{i1}h_{i2}(h_{i1}+h_{i2})}.
      \end{equation}

\begin{figure}
	\includegraphics[width=\textwidth]{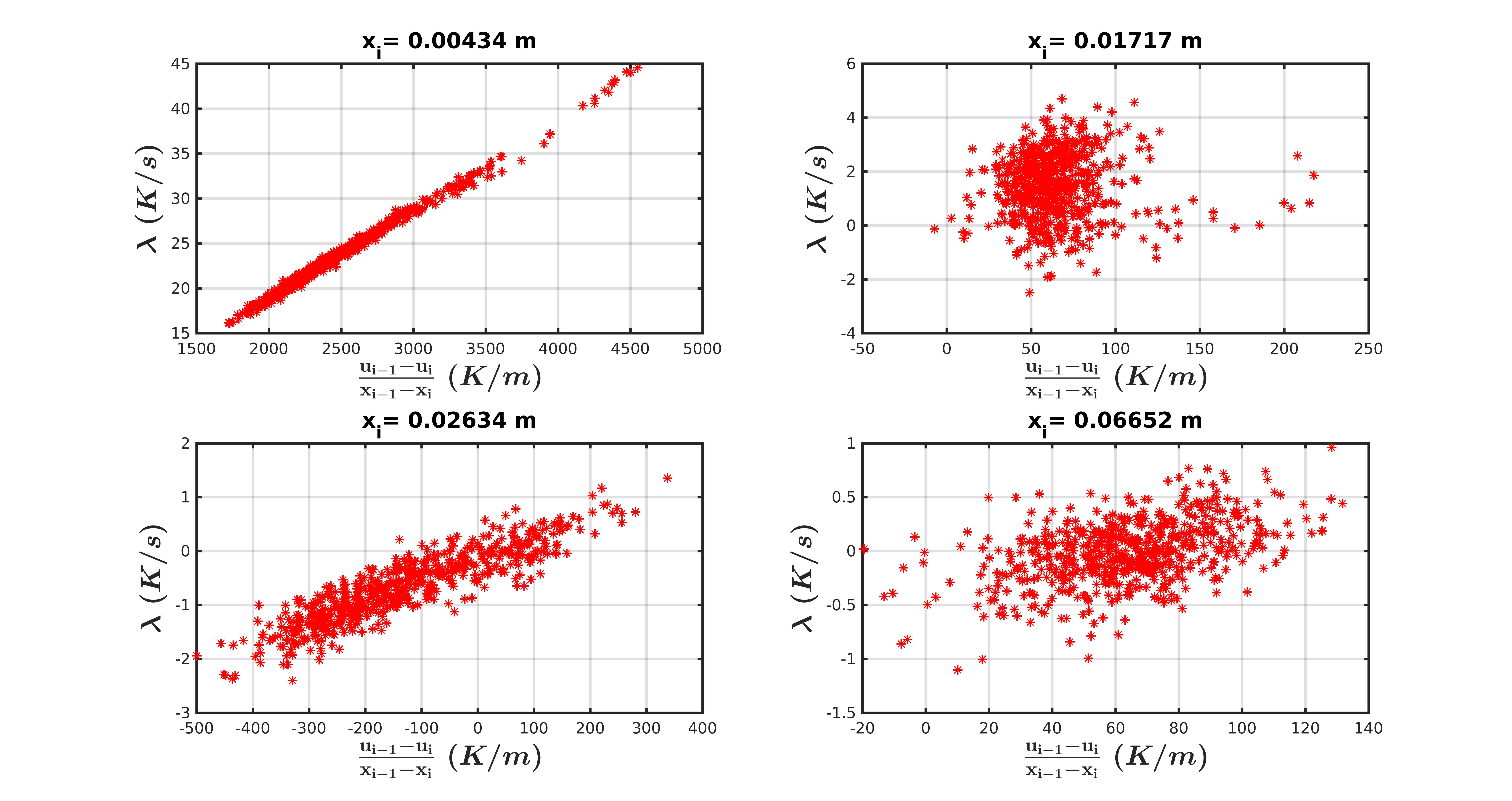}
	\includegraphics[width=\textwidth]{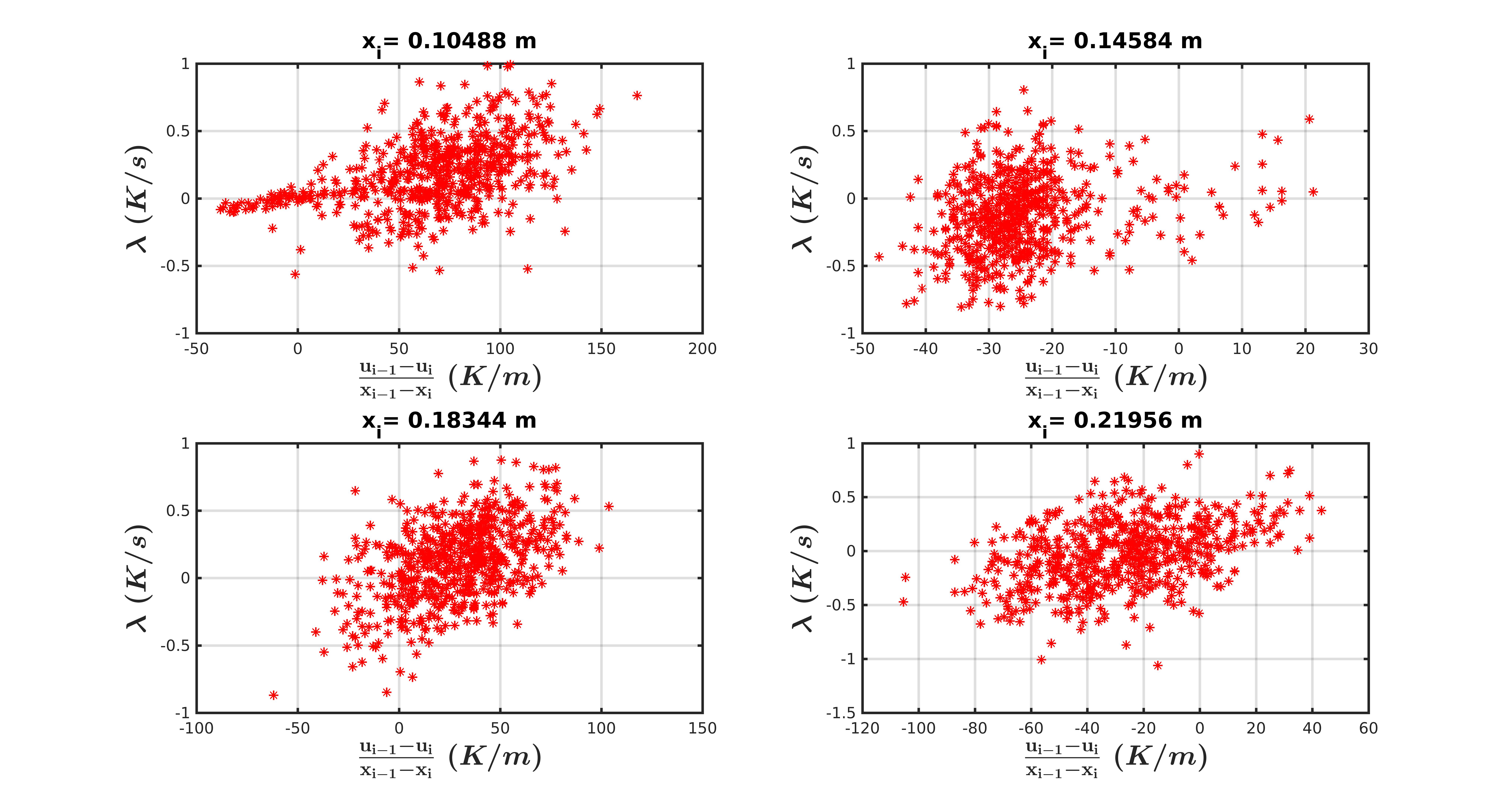}
	\includegraphics[trim={0 70cm 0 0},clip,width=\textwidth,height=3.8cm,angle=0]{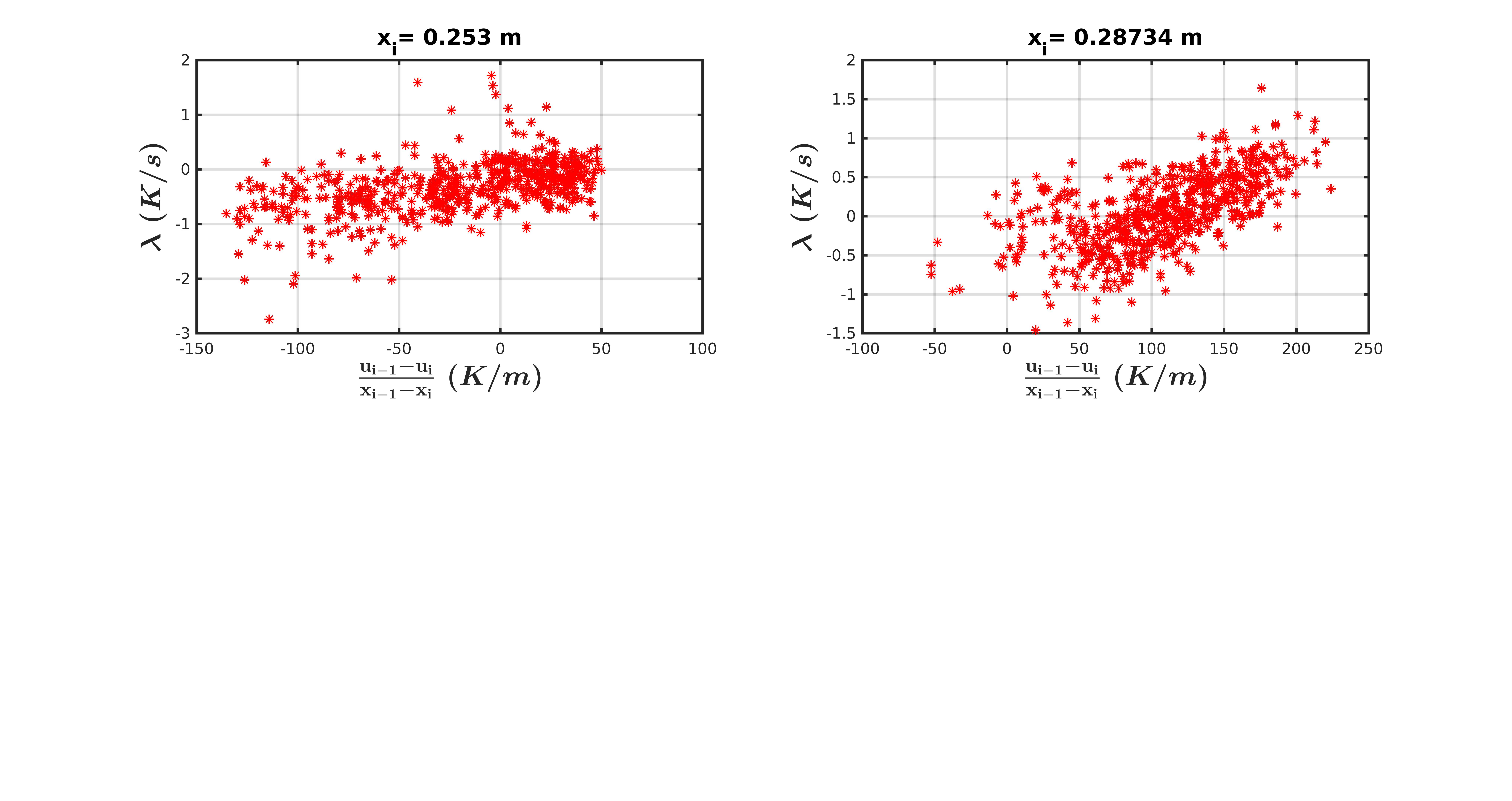}
	\caption{Lagrange multiplier $\lambda$ plotted against approximate gradient of temperature $\frac{u_{i-1}-u_{i}}{x_{i-1}-x_i}:=D_i(u_i)\approx \frac{\partial u}{\partial x}$ shows a linear variation.} \label{fig7}
\end{figure}

\begin{figure}
	\includegraphics[width=\textwidth]{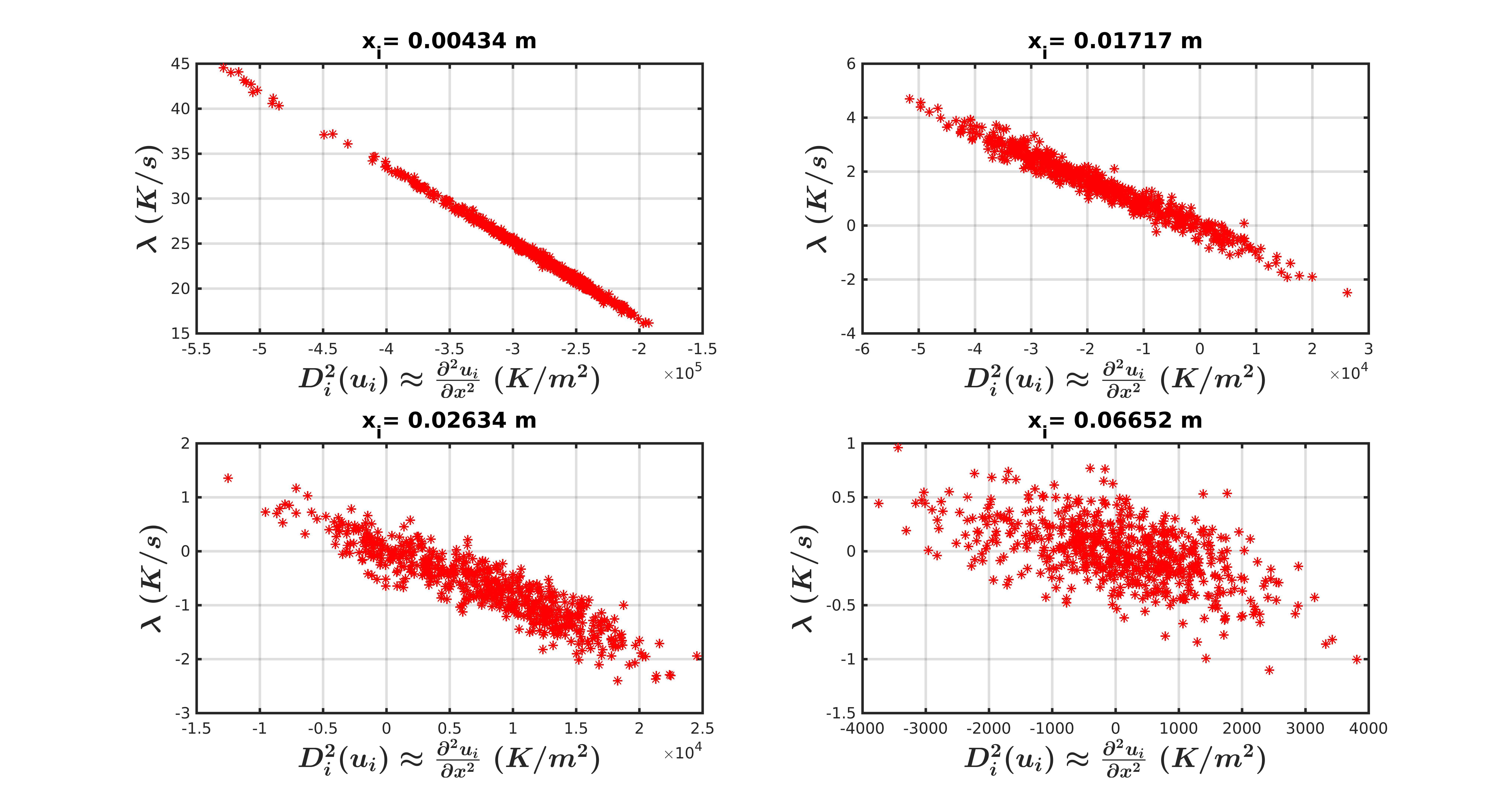}
	\includegraphics[width=\textwidth]{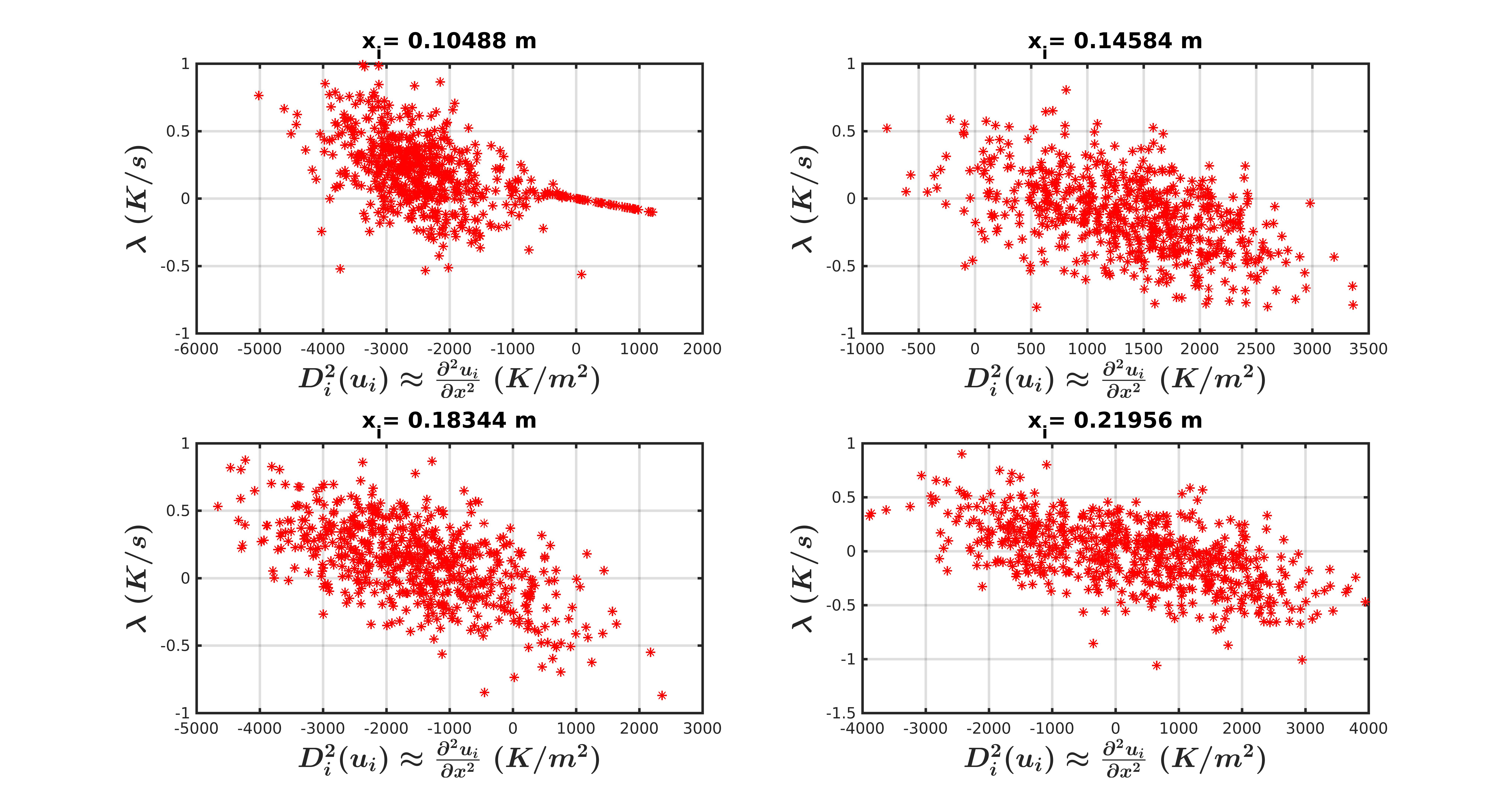}
	\includegraphics[trim={0 70cm 0 0},clip,width=\textwidth,height=3.8cm,angle=0]{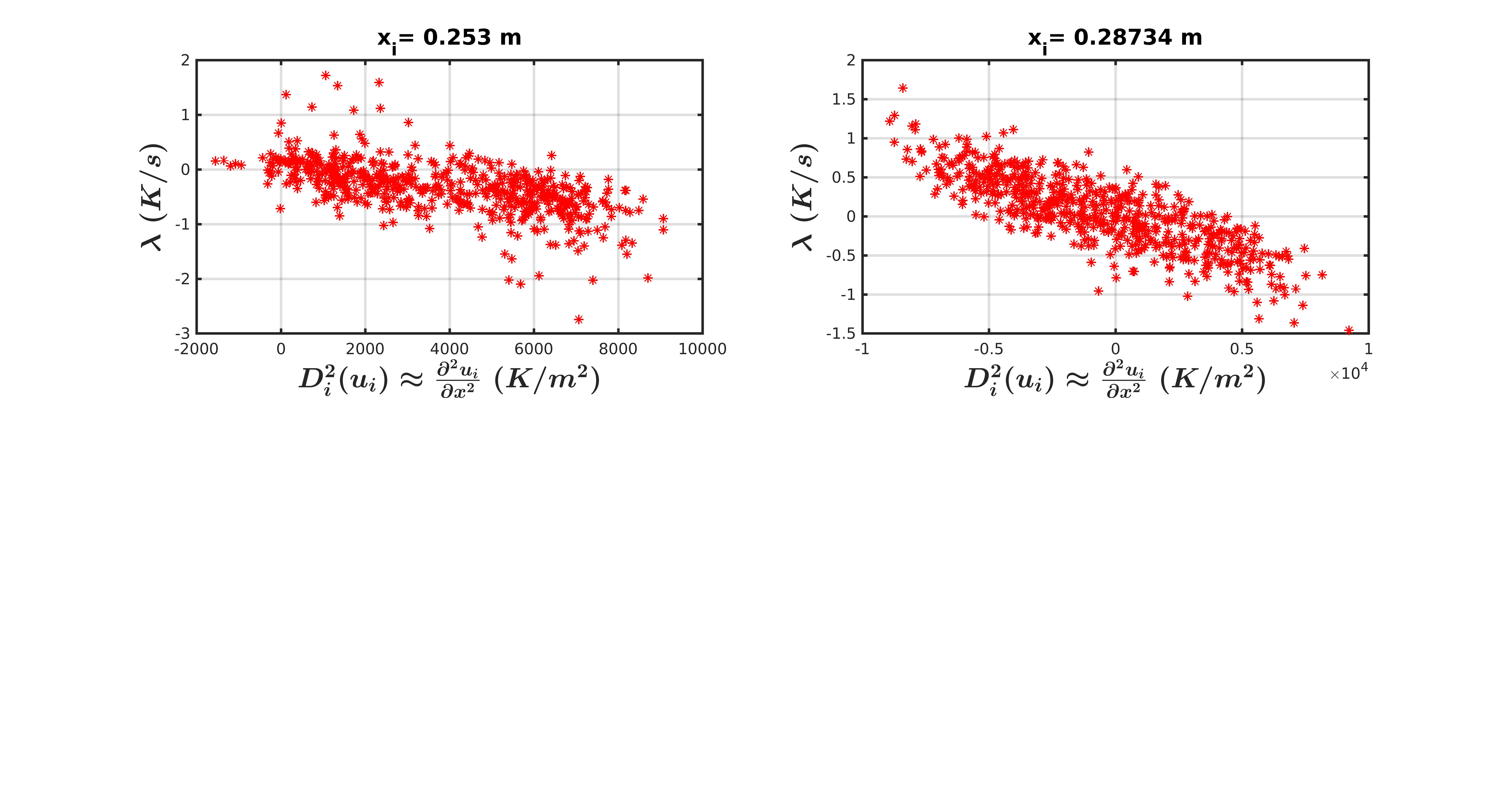}
	\caption{For all nodes, linear trend is also observed when Lagrange multiplier $\lambda$ (at the $i^{th}$ node) is plotted against finite difference approximation of second spatial derivative of u represented by $D_i^2{u_i}:=2\frac{h_{i1}u_{i-1}-(h_{i1}+h_{i2})u_{i}+h_{i2}u_{i+1}}{h_{i1}h_{i2}(h_{i1}+h_{i2})} \approx \frac{\partial^2 u_i}{\partial x^2}$.} \label{fig8}
\end{figure}

\begin{figure}[t]
\centering
	\begin{subfigure}{0.5\textwidth}
		\includegraphics[width=\textwidth]{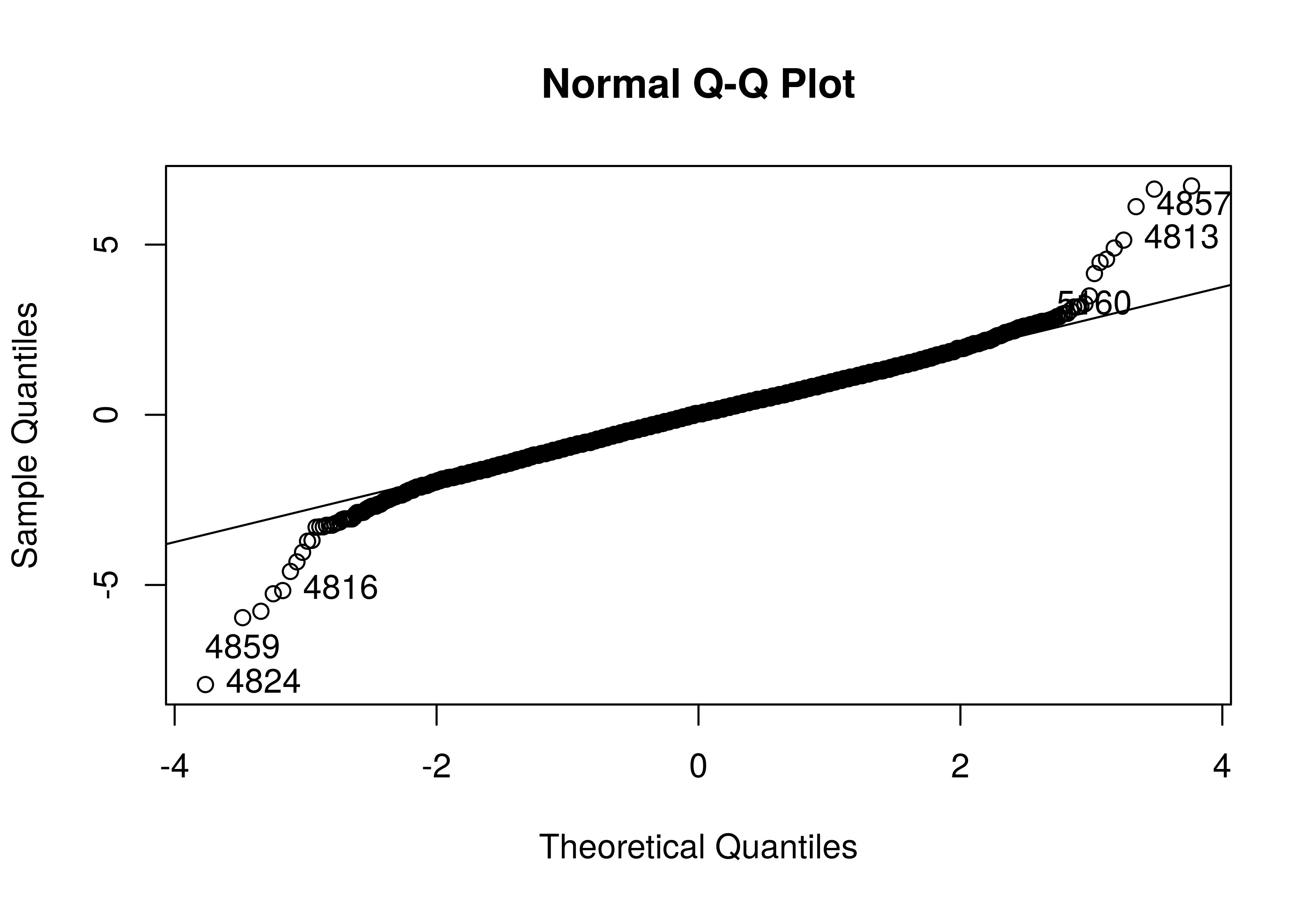}
		\subcaption{} 
	\end{subfigure}%
	\begin{subfigure}{0.5\textwidth}
		\includegraphics[width=\textwidth]{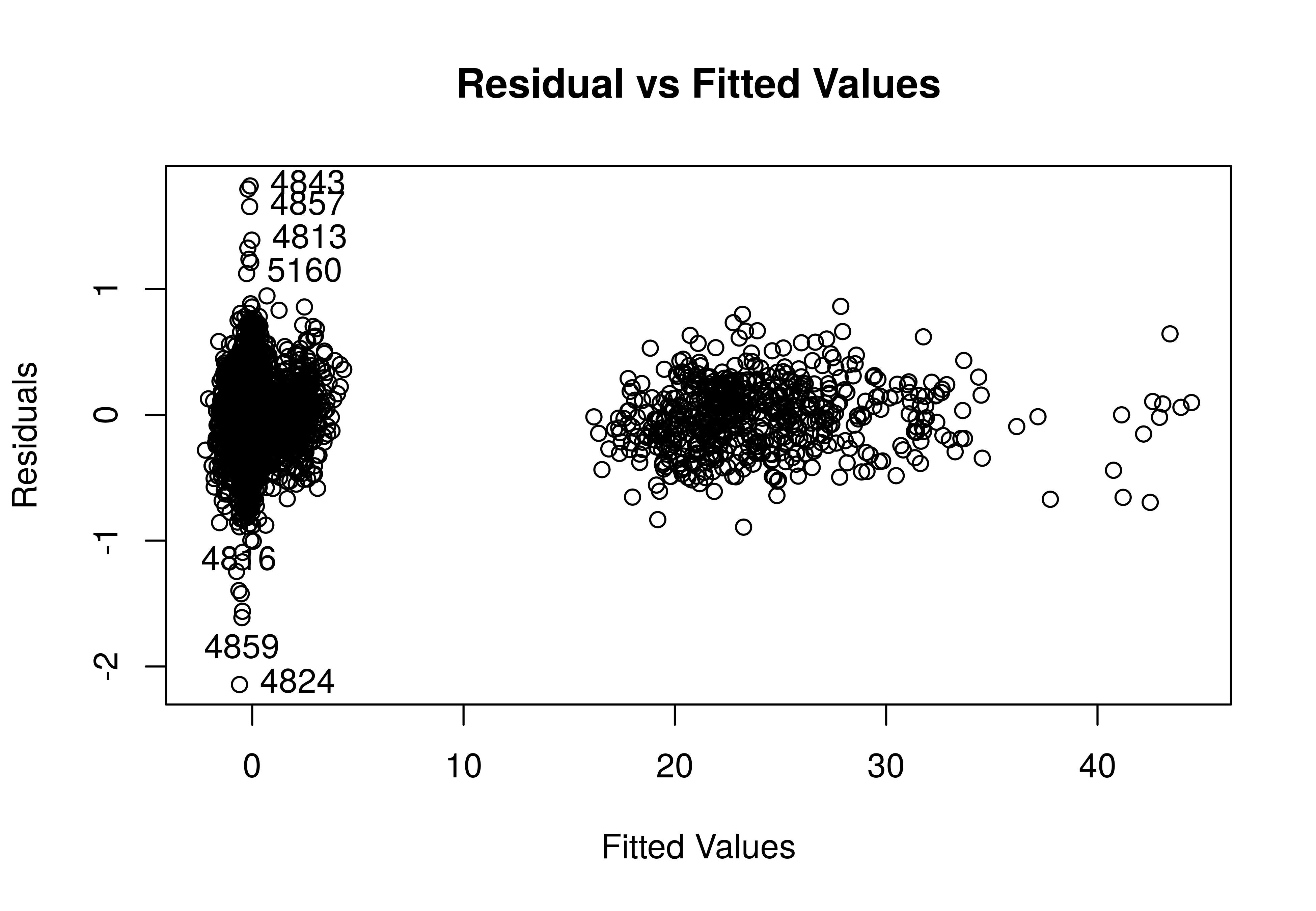}
		\subcaption{} 
	\end{subfigure}
	\begin{subfigure}{0.5\textwidth}
		\includegraphics[width=\textwidth]{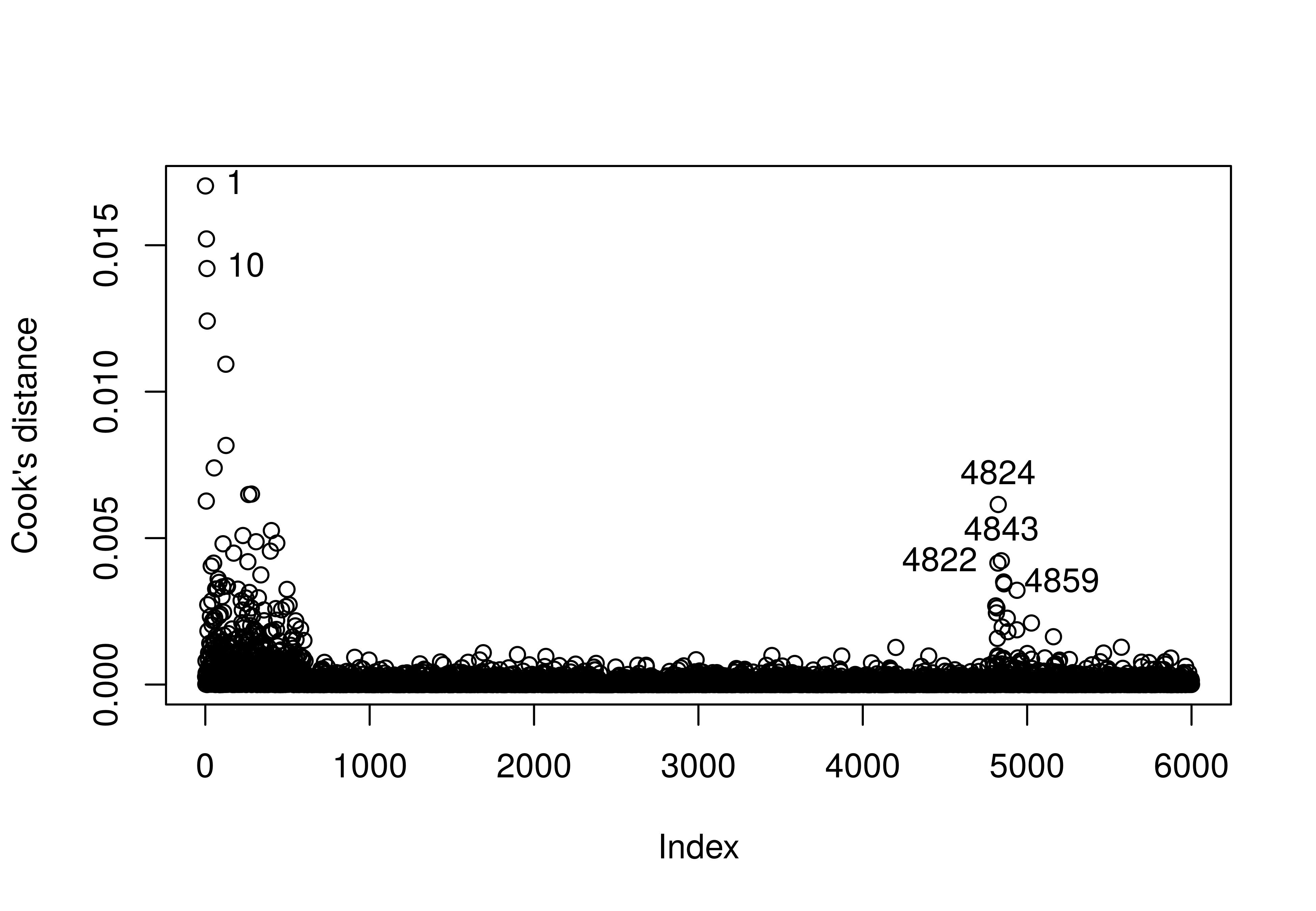}
		\subcaption{} \label{fig17}
	\end{subfigure}
\caption{(a) Normal Probability plot of standardized residuals for regression model that fits $\lambda$ values computed from the first partition data with the corresponding second spatial derivative of u, (b) Residuals vs fitted values plot for this fit, (c) Cook's distance of $\lambda$ values indexed with Integers. Some of the common data points are indicated by their indexes against them in all plots.}\label{fig9}
\end{figure}

If $\frac{\partial^2 u_i}{\partial x^2} =\psi \lambda_i+c_3$, $\frac{\partial u_{i}}{\partial x}=\gamma \lambda_i+c_2$ and $u_{i-1}=c_1$, for some constants 
$\psi$, $\gamma$, $c_1$, $c_2$ and $c_3$, then from \ref{36}, we have,
      \begin{equation}
           \psi \lambda_i+c_3  \approx 2\frac{h_{i1}c_1-(h_{i1}+h_{i2})u_{i}+h_{i2}u_{i+1}}{h_{i1}h_{i2}(h_{i1}+h_{i2})}
      \end{equation}
      \begin{align}
 \implies     \psi \lambda_i-2\frac{h_{i1}c_1}{h_{i1}h_{i2}(h_{i1}+h_{i2})}+c_3 &\approx 2\frac{h_{i2}(u_{i+1}-u_i)-h_{i1}u_i}{h_{i1}h_{i2}(h_{i1}+h_{i2})}\\
\implies \psi \lambda_i-2\frac{h_{i1}c_1}{h_{i1}h_{i2}(h_{i1}+h_{i2})}+c_3 &\approx 2\frac{\gamma \lambda_i+c_2}{h_{i1}+h_{i2}}-2\frac{h_{i1}u_i}{h_{i1}h_{i2}(h_{i1}+h_{i2})}\\
\implies u_i \approx h_{i2}\gamma \lambda_i -\frac{1}{2}\psi \lambda_i h_{i2}(h_{i1}+&h_{i2})+h_{i2}c_2+c_1-c_3\frac{h_{i2}}{2}(h_{i1}+h_{i2}).
      \end{align}
So, if we assume a linear relationship between $\lambda$ and first and second partial derivatives of $u$ with respect to $x$, then we can expect a linear relationship between $u_1$ and $\lambda_1$.\\
 The above discussion motivates us to check for a linear relationship between $\lambda$ and first and second partial derivatives of u with respect to x.
Define $D_i(u_i)=\frac{u_i-u_{i-1}}{x_i-x_{i-1}}\approx \frac{\partial u_i}{\partial x}$ and $D^2_i(u_i)=2\frac{h_{i1}u_{i-1}-(h_{i1}+h_{i2})u_{i}+h_{i2}u_{i+1}}{h_{i1}h_{i2}(h_{i1}+h_{i2})}\approx  \frac{\partial^2 u_i}{\partial x^2}$.
Figures \ref{fig7} and \ref{fig8} depict the plots of $\lambda_i$ with $D_i(u_i)$ and $D^2_i(u_i)$ respectively. These plots do indicate a linear variation between these variables. 
We find suitable multiple regression model to estimate $\lambda_i$'s with regressors variables being  elements of non-empty subsets of $\{u_i,D_i(u_i),D^2_i(u_i)\}$.
Since the available dataset is small, $\lambda_i$'s from all nodes with the corresponding regressor variables are put together and indexed with integer values; and a single regression model is fitted to estimate $\lambda$. A small note about the notation, $\lambda_i$ denotes the Lagrange multiplier at the measurement node $x_i$. However, we are fitting one single regression model for $\lambda_i$'s taken from all nodes. Hence the estimated Lagrange multiplier will not depend on the node but just on the regressor variable computed at the node. So, we will use $\lambda_i$ or $\lambda$ according to the context. The $R^2$ and adjusted $R^2$ values for models with different sets of regressor variables in this case is tabulated in Table \ref{tab2}. With only $D^2_i(u_i)$ as regressor variable, the coefficient of determination $R^2$ is 0.9986527. Adding more regressor variables doesn't change the coefficient of determination much. Similar trend is also seen for adjusted $R^2$. Hence, a simple linear 
regression model consisiting of only one regressor $D^2_i(u_i)$ is fitted to estimate $\lambda$. References \cite{Montgomery} and \cite{Draper} are good sources for detailed discussion on regression analysis.
\begin{table}[!ht] 
\caption{$R^2$ and adjusted $R^2$ for different combination of regressor variables.} \label{tab2}
 \centering
    \begin{tabular}{|c| c|c|}
    \hline
    Regressors & $R^2$ & Adjusted $R^2$ \\
     \hline
  $u$ &  0.07009437 & 0.06993933\\
  \hline
   $D_i(u_i)$ & 0.9890581 & 0.9890563 \\
  \hline
  $D_i^2(u_i)$ & 0.9986527 & 0.9986525  \\
    \hline
    $u$,$D_i(u_i)$ & 0.9909068 & 0.9909038  \\
    \hline
    $u$,$D_i^2(u_i)$ & 0.998661 & 0.9986605 \\ 
    \hline
    $D_i(u_i)$,$D_i^2(u_i)$ & 0.9986528 & 0.9986524 \\
    \hline
   $u$, $D_i(u_i)$,$D_i^2(u_i)$ & 0.9986622 & 0.9986615\\
   \hline
    \end{tabular}
\end{table}
Normal probability plot of the standardized residuals for this regression fit is shown in figure \ref{fig9}a. 
Few points($\approx 0.3\%$) have large deviation from straight line in the normal probability plot. The residuals corresponding to these points also show larger spread compared to other residuals when plotted against the fitted values as shown in figure \ref{fig9}b.
Some of the common data indexes between figures \ref{fig9}a, \ref{fig9}b and \ref{fig9}c are highlighted.\par
\begin{figure}
	\centering
		\begin{subfigure}{0.5\textwidth}
			\includegraphics[width=\textwidth]{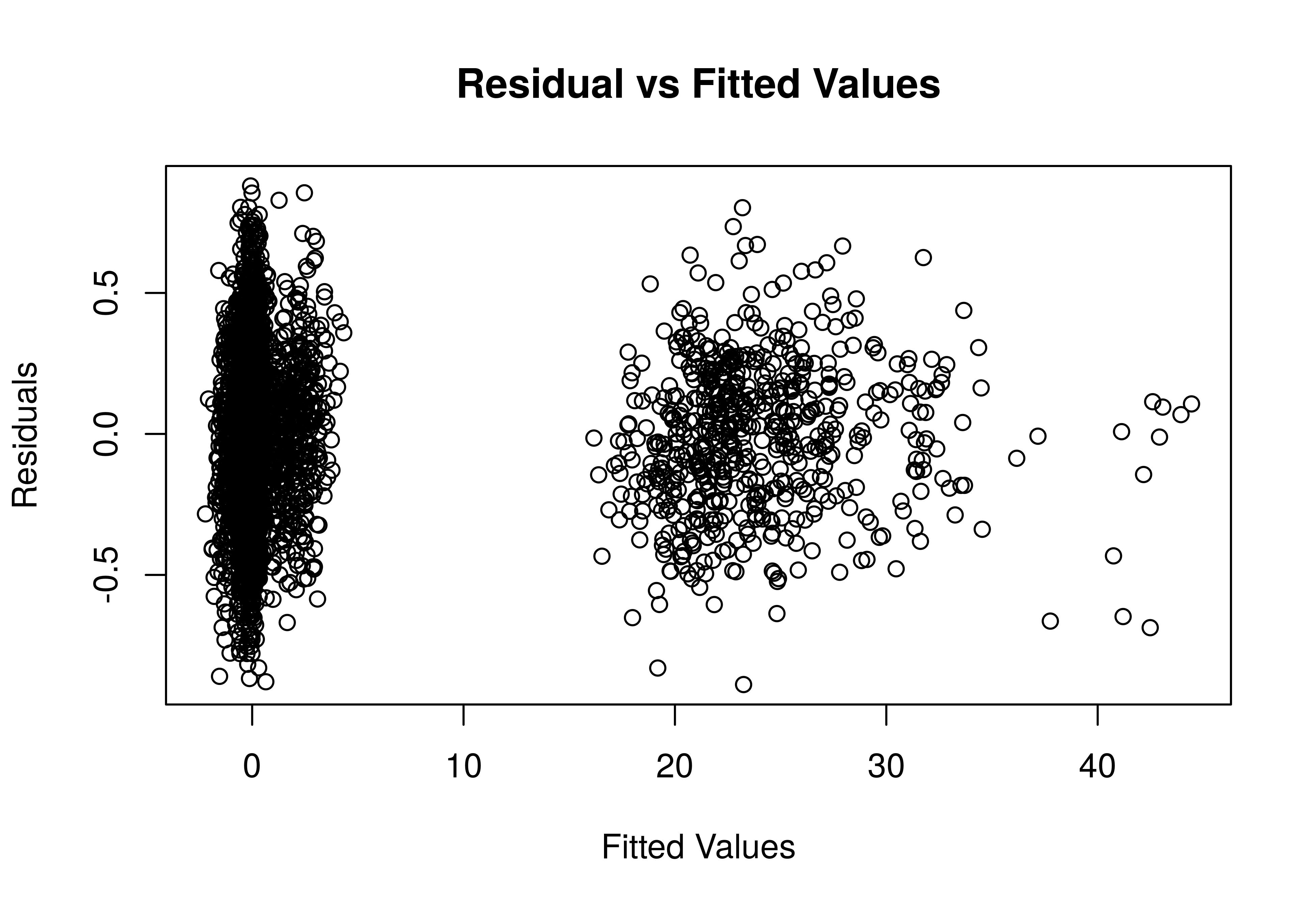}
			\subcaption{}
		\end{subfigure}%
		\begin{subfigure}{0.5\textwidth}
			\includegraphics[width=\textwidth]{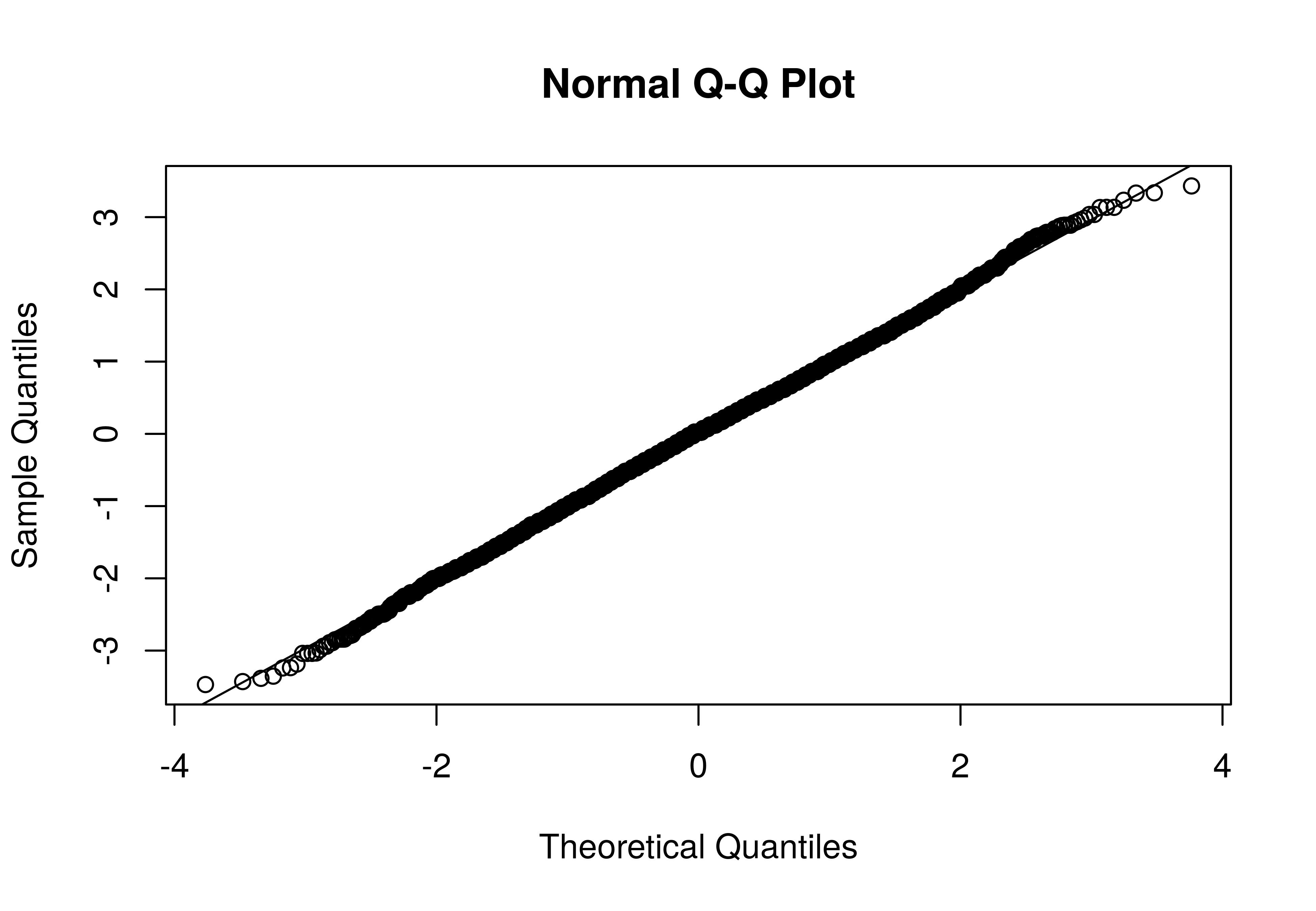}
			\subcaption{}
		\end{subfigure}
	\caption{(a) Residuals vs fitted values, (b) Normal probability plot for the new regression model.} \label{fig10}
\end{figure}

Figure \ref{fig9}c shows the Cook's distance for all the values of $\lambda$. The points departing from the model assumptions are also
influential observations. The effect of removing these possible outliers is inspected next.\newline
Temperatures are predicted using a regression model for $\lambda$ that is fitted through all datapoints and the prediction performance is compared with another model whose coefficients are computed after removing the influential observations that do not follow the model assumptions. 
Estimated values of $\lambda_i$ from either of the two fitted regression models is given by the following expression:
\begin{equation}
 \hat{\lambda}_i=\hat{\beta_0}+\hat{\beta_1}D^2_i(u_i). \label{41}
\end{equation}
Here, $\hat{\beta_0}$ and $\hat{\beta_1}$ are regression coefficients. These regression coefficients assume different values for the two regression models---one fitted with all available $\lambda_i$ values and another by omitting the possible outliers.
The following equations are then solved to predict temperatures at each node by additing an extra $\hat{\lambda}_i$ term, obtained from regression model \ref{41}, to the nominal heat conduction model.
\begin{align} \label{42}
\frac{\partial \hat{u}_i}{\partial t}=\frac{\partial^2 \hat{u}_i}{\partial x^2}+\hat{\beta}_1D_i^2(u_i)+\hat{\beta}_0,
\end{align}
\begin{align} \label{43}
\frac{\partial \hat{u}_i}{\partial t}=\frac{\partial^2 \hat{u}_i}{\partial x^2}+\hat{\beta}_1D_i^2(\hat{u}_i)+\hat{\beta}_0.
\end{align}
Equations \ref{42} and \ref{43} are used to obtain two different estimates $\hat{u}_i$ of temperature at the $i^{th}$ node. Both of these equations are similar to equation \ref{25a} except that $\lambda_i$ is replaced by the estimate $\hat{\lambda}_i$ from the regression model \ref{41}. It is to be noted 
that in \ref{42}, the  independent variable used in the fitted regression model to compute $\hat{\lambda}_i$ is $D_i^2(u_i)$ which is a function of the 
observed temperature $u_i$. But in \ref{43}, the estimate  $\hat{\lambda}_i$ is obtained using $D_i^2(\hat{u}_i)$ which is the finite difference
approximation of second derivative of the temperature estimate $\hat{u}_i$.
Equations \ref{42} and \ref{43} are solved with two different sets of $(\beta_0,\beta_1)$ obtained using regression fit with possible outliers included and removed respectively. Mean squared error(MSE) between the observed and estimated temperatures for each case is tabulated in table \ref{tab3}.
\begin{table}
 \centering
\caption{MSE for estimated values of temperature obtained as solutions of \ref{42} and \ref{43} with respect to its observed values when 1) Regression model for $\lambda$ is fitted using all data, 2) Regression model for $\lambda$ is fitted after removing suspected outliers from the dataset.} \label{tab3}
    \begin{tabular}{c c|c|} 
\cline{2-3}
& \multicolumn{2}{|c|}{MSE $(K^2)$}\\
\cline{2-3}
 &  \multicolumn{1}{|c}Eq \ref{42} & Eq \ref{43} \\
  \hline
  \multicolumn{1}{|c|}{Regression coefficients $\beta_0$, $\beta_1$ determined using all datapoints} & 0.355 &  0.986\\
     \hline
  \multicolumn{1}{|c|}{Regression coefficients $\beta_0$, $\beta_1$ determined after removing}  &  0.323 & 0.978\\
    \multicolumn{1}{|c|}{ influential observations that violate model assumptions}  & & \\
\hline
    \end{tabular}
\end{table}
We observe that the MSE reduces when datapoints that violate the regression assumptions are removed from the dataset. This reduction in MSE is observed for estimates obtained from both \ref{42} and \ref{43}. 
These small number of influential observations which violate the regression assumptions are excluded and coefficients $\hat{\beta}_0$ and $\hat{\beta}_1$ are estimated using the remaining data points. Diagnostic plots for the new regression fit are shown in figure \ref{fig10}. It can be observed that the residuals in this fit are roughly normally distributed with constant variance. 
 \\
The model for $\lambda_i$ for $i\in\{1,2,\dots,n-1\}$ is thus expressed as:
\begin{align}
\lambda_i=\beta_0+\beta_1D_i^2(u_i)+\epsilon_i, 
\end{align}
where $\epsilon_i \stackrel{iid}{\sim} \mathcal{N}(0,\,\sigma^{2})$. $\hat{\beta}_0$ and $\hat{\beta}_1$ are the estimator of $\beta_0$ and $\beta_1$. \\
Denoting the residual sum of squares by $SS_{Res}$, the unbiased estimator of $\sigma^2$ is given by:
\begin{align}
\hat{\sigma}^2=\frac{SS_{Res}}{N-K}, \label{45}
\end{align}
where N is the number of observations for fitting the regression model and K is the number of regressor variables plus one. In the present model, $N=5968$ and $K=2$. The analysis presented till here involves only the first partition of the dataset. Now we make predicitons and compare it with the second partition 
of the measurement data.\\
Prediction equations are formed by including Lagrange multiplier term in the nominal heat equation. This term is estimated from the regression model described in the preceding discussion. Therefore, the prediction equation is given as follows:
\begin{align}
 \frac{\partial u_i}{\partial t}=\alpha\frac{\partial ^2 u_i}{\partial x^2}+\hat{\lambda}_i. \label{46}
\end{align}
We continue with the same notations for mesh points and corresponding temperature values as in equation \ref{30.1}. By backward time centered space finite difference approximation of \ref{46}, we get,
\begin{equation} \label{47}
 \frac{u_{i}^k-u_{i}^{k-1}}{\Delta t}=2\alpha\frac{h_{i1}u_{i-1}^{k}-(h_{i1}+h_{i2})u_{i}^k+h_{i2}u_{i+1}^k}{h_{i1}h_{i2}(h_{i1}+h_{i2})}+\hat{\lambda}_{i}^k.
\end{equation}
Replacing $D^2_i(u_i)$ with its finite difference approximation in equation \ref{41} and rewriting it for the $k^{th}$ time step we get,
\begin{align} \label{48}
\hat{\lambda}_i^k=&\hat{\beta}_0+2\hat{\beta}_1 \frac{h_{i1}u_{i-1}^{k}-(h_{i1}+h_{i2})u_{i}^k+h_{i2}u_{i+1}^k}{h_{i1}h_{i2}(h_{i1}+h_{i2})}.
\end{align}
Substituting $\hat{\lambda}_i^k$ from \ref{48} in \ref{47} ,
 \begin{equation} \label{49}
   \frac{u_{i}^k-u_i^{k-1}}{\Delta t}=2(\alpha+\hat{\beta}_1)\frac{h_{i1}u_{i-1}^k-(h_{i1}+h_{i2})u_i^k+h_{i2}u_{i+1}^k}{h_{i1}h_{i2}(h_{i1}+h_{i2})}
   +\hat{\beta}_{0}.
\end{equation}
Equation \ref{49} is similar to equation \ref{30.1} except that $\alpha$ and $\lambda_i^k$ are replaced by $\alpha+\hat{\beta}_1$  and $\hat{\beta}_0$ respectively. Therefore, on putting together a single vector equation for all i's, we arrive at an equation similar to equation \ref{32}. It is given as:
\begin{align}
\boldsymbol{Au^k}=\boldsymbol{u^{k-1}}+\Delta t \boldsymbol{\hat{\beta}_0} \label{50}, 
\end{align}
where,
\begin{equation*}
{\boldsymbol{A}=
\begin{bmatrix}
         1 & 0 & \dots & \dots & \dots & 0\\
         a_1 & b_1 & c_1 & 0 & \dots & 0 \\
	 0 &  a_2 & b_2 & c_2  & \dots & 0 \\
	 \vdots & \vdots & \vdots & \vdots & \vdots & \vdots \\
	 \vdots & \vdots & \vdots & \vdots & \vdots & \vdots\\
	 0 & 0 & \dots &  a_{n-1} & b_{n-1} & c_{n-1} \\
	 0 & 0 & \dots & 0 & 0 & 1
	 
   \end{bmatrix}
   }
\end{equation*}with
\begin{align}
a_i=&-\frac{2\Delta t(\alpha +\hat{\beta}_1)}{h_{i2}(h_{i1}+h_{i2})}\\
b_i=&1+\frac{2\Delta t(\alpha+\hat{\beta}_1)}{h_{i1}h_{i2}}\\
c_i=&-\frac{2\Delta t(\alpha+\hat{\beta}_1)}{h_{i1}(h_{i1}+h_{i2})},
\end{align}
$\boldsymbol{\hat{\beta}_0}$ is an  $(n+1)$-vector with all elements being the estimated regression coefficient $\hat{\beta}_0$. 
 \begin{figure}
\centering
    \includegraphics[width=10.5cm,height=6.5cm,angle=0]{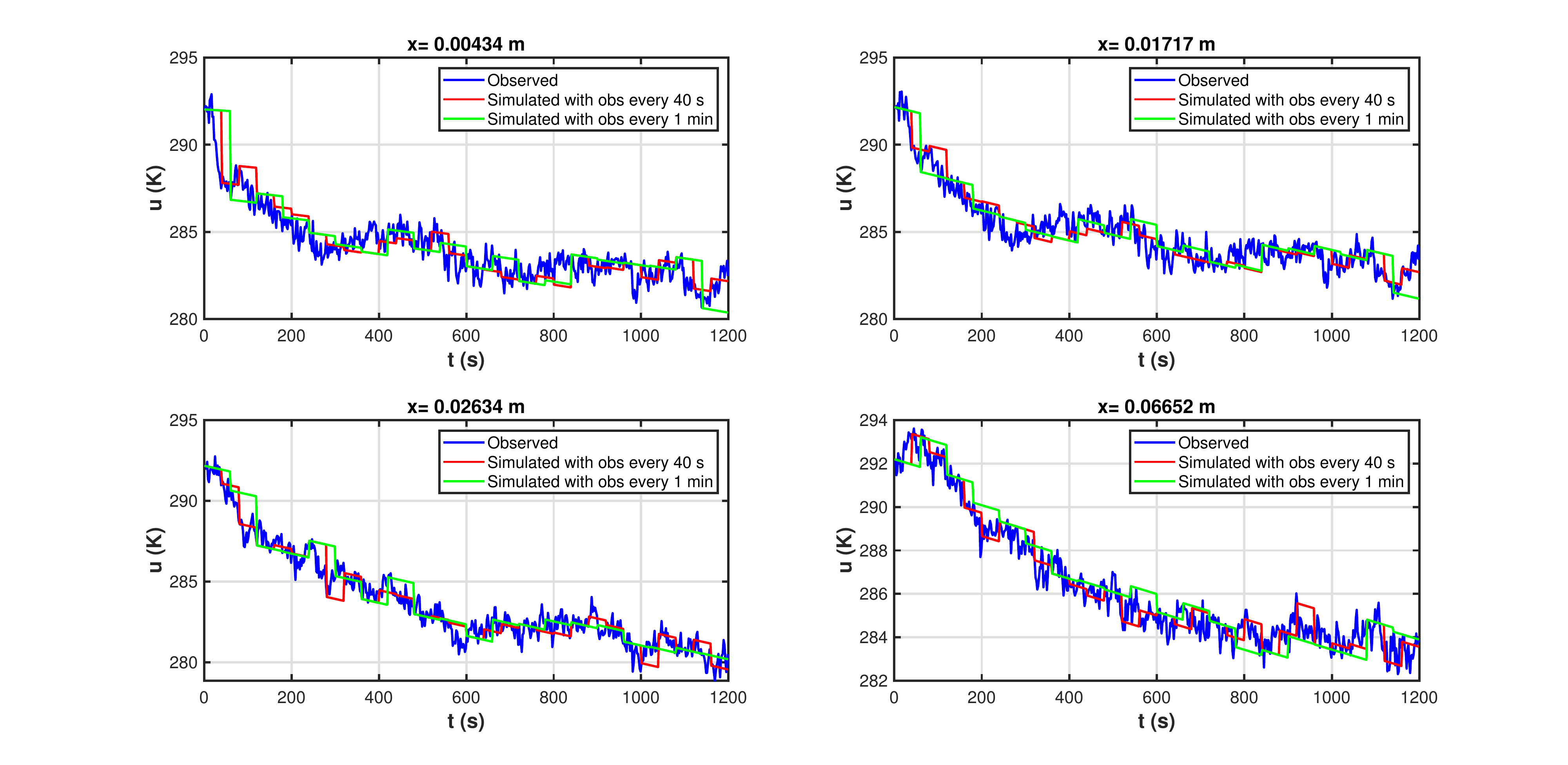}
    \includegraphics[width=10.5cm,height=6.5cm,angle=0]{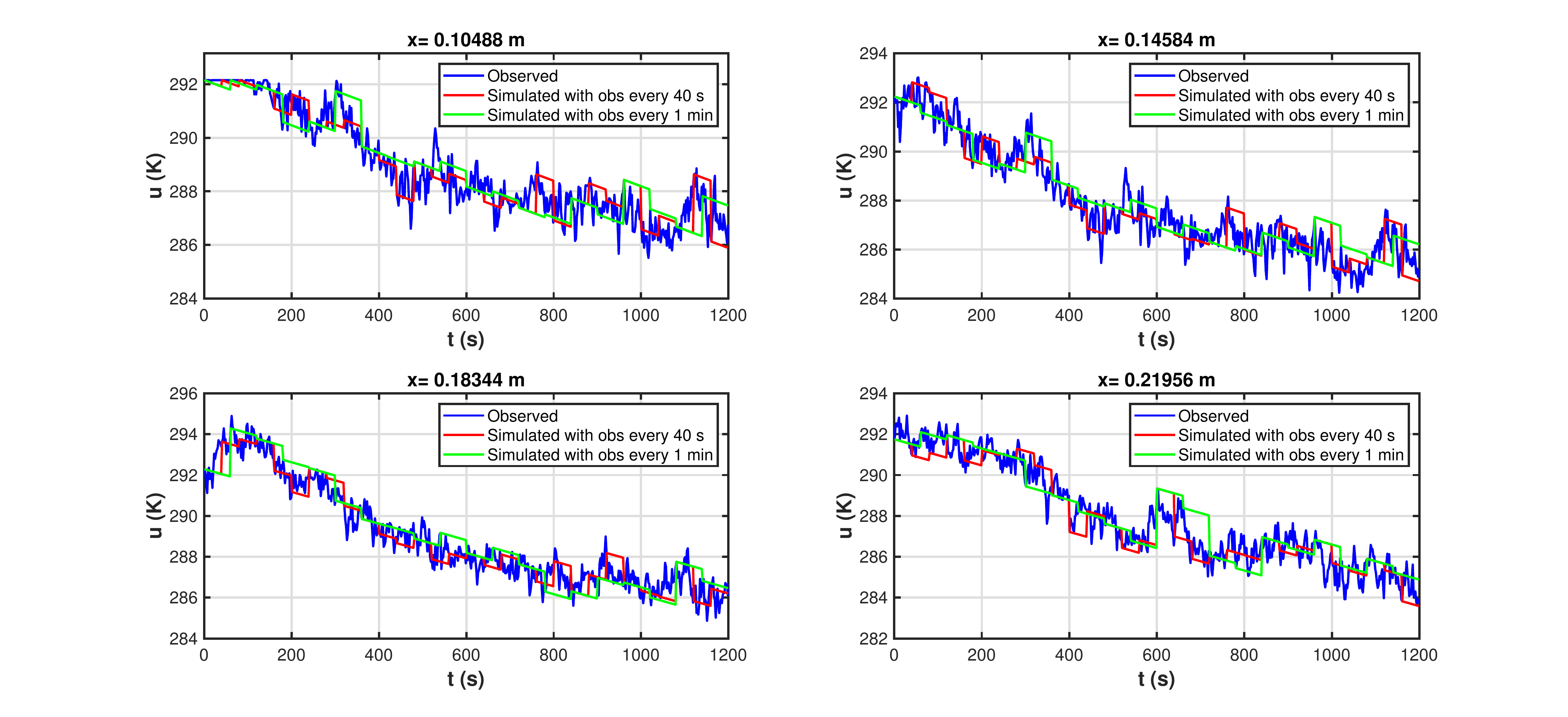}
    \includegraphics[trim={0 8.9cm 0 0},clip,width=10.5cm,height=3.7cm,angle=0]{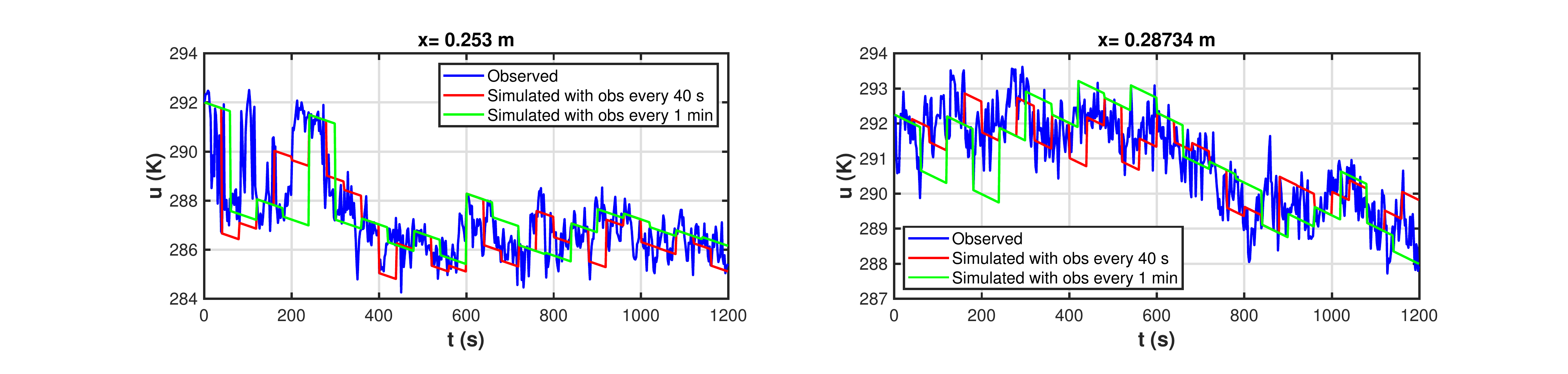}
    \caption{Comparision between the observed temperature (blue) and modified heat equation solution initialized with observations every \SI{40}{s} (red) and \SI{1}{min} (green). Nominal heat equation is modified with Lagrange multiplier term estimated from the regression model. Before testing on second partition test dataset, the modified model performance is evaluated in predicting the observations in the first data partition.} \label{train}
\end{figure}
\begin{figure}
	\centering
	\includegraphics[width=10.5cm,height=6.5cm,angle=0]{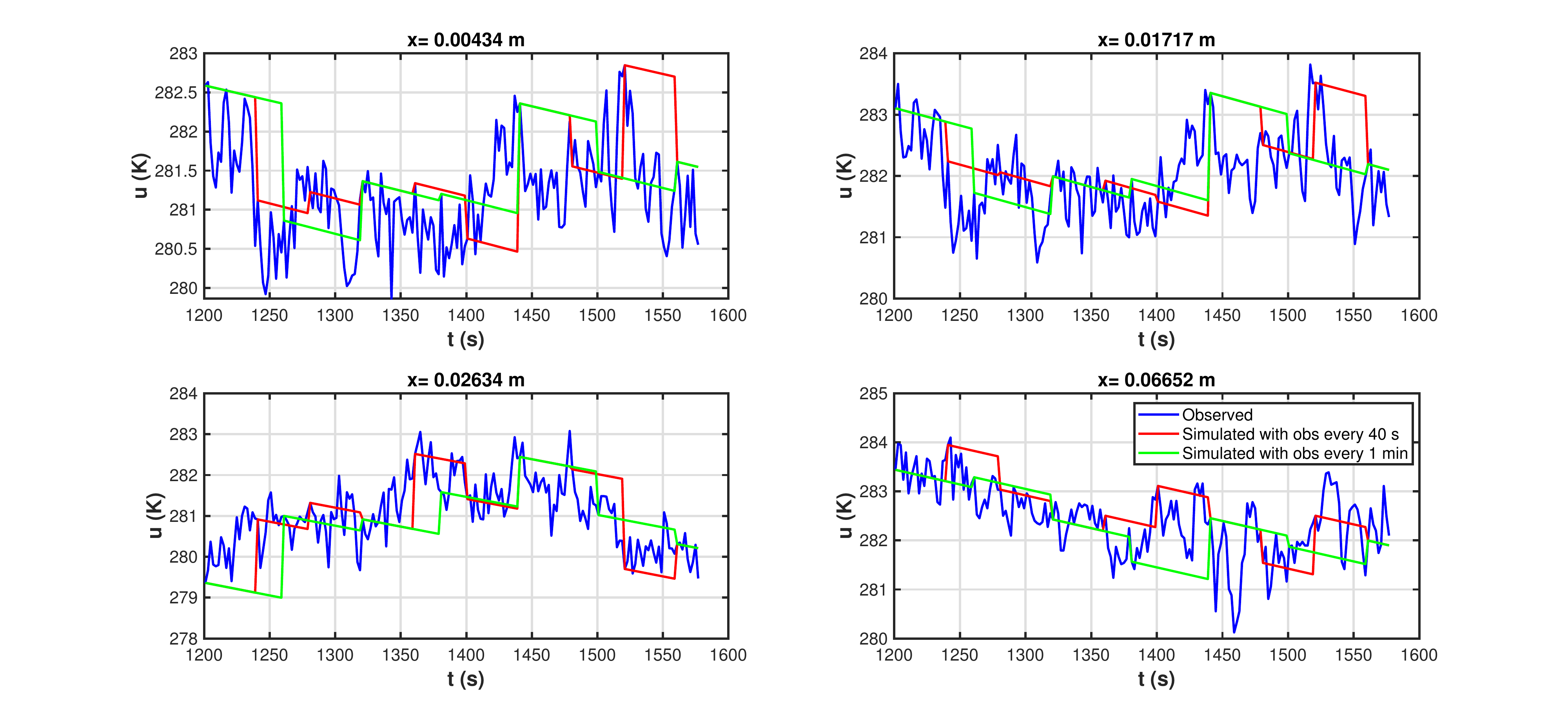}
	\includegraphics[width=10.5cm,height=6.5cm,angle=0]{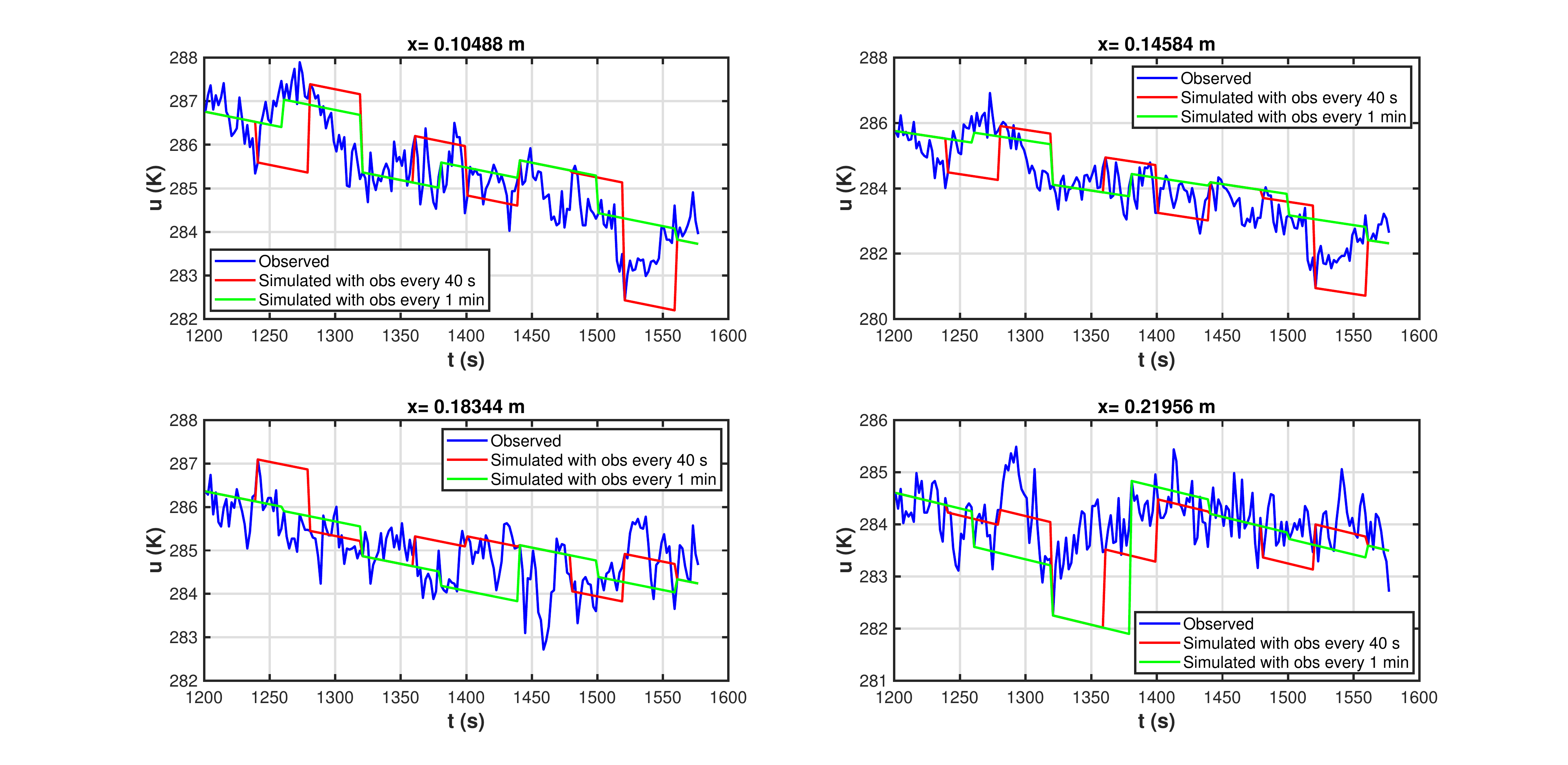}
	\includegraphics[trim={0 8.9cm 0 0},clip,width=10.5cm,height=3.7cm,angle=0]{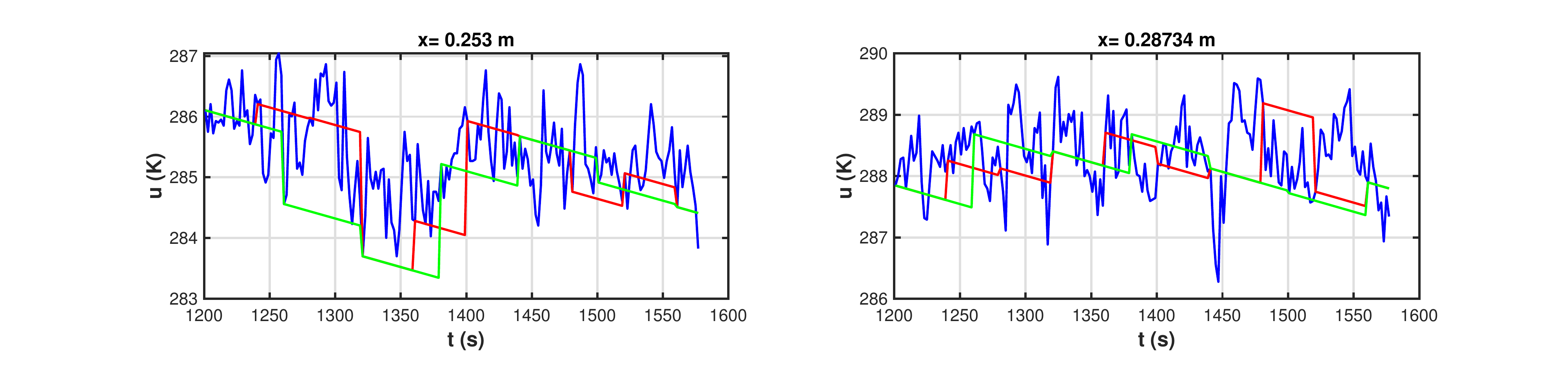}
	\caption{Comparision between the observed temperature (blue) and modified heat equation solution initialized with observations every \SI{40}{s} (red) and \SI{1}{min} (green) for the second data partition. Observed temperatures from the first data partition are used to modify the nominal heat equation. Performance of the modified model is tested with respect to the second data partition and the predicted temperatures are presented here.} \label{test}
\end{figure}
\par
Before evaluating the prediction performance of the modified model with respect to the second partition test data, we check its performance for
predicting the observations in the first data partition. Essentially to check the performance of the new model equations, we are attempting to predict back the data which was used to modify the model. For this, we solve equation \ref{50} with $\boldsymbol u$ intialized with the first temperature observation at t=0.9 in the first data partition. As the accuracy of our analysis is limited by a small training set of 600 datapoints only, we predict for a short period of \SI{38}{s} and \SI{58}{s} and reinitialize equation \ref{50} with the available data on the next time step. In this way, we are able to test the performance for several initial conditions and sample paths. Figure \ref{train} shows a comparison of the observed temperature with the 
predicted temperature when the prediction equation is reintialized every \SI{40}{s} and \SI{60}{s}. A steep decline in the temperature in comparison to the observed temperature was seen when the nominal heat equation model was solved as depicted in figure \ref{fig1a}. Solutions from the modified model equations shown in figure \ref{train} seem to perform better in this regard. Performance measure is computed in terms of Mean squared Error(MSE) between the observed and predicted temperatures, taking together the data for all measurement nodes and time steps. MSE is $\SI{0.70}{K^2}$, when \ref{50} is reinitialized every \SI{40}{s}. When initialization is done after every \SI{60}{s}, MSE is $\SI{0.98}{K^2}$. In comparison, predicted temperatures obtained as solutions of nominal heat equation initialized with observations every \SI{40}{s} and \SI{60}{s} have MSE of \SI{21.55}{K^2} and \SI{25.28}{K^2} respectively with respect to their observed values. \par
Now we evaluate the modified model's ability to predict the second partition of the dataset which is not used in any previous analysis. The first partition consisted of the first 600 data points. The next observation i.e. $601^{th}$ datapoint is recorded at \SI{1200.9}{s} and is the first observation of the second partition dataset. Equation \ref{50} is solved  with this observation as the initial condition to get the value of $\mathbf{u}(t)$ for t in range of \SI{1202.9}{s} to \SI{1576.9}{s}. As previously, we reinitialize with observations every \SI{40}{s} and \SI{60}{s}, so that we are predicting against several test sets for a short time interval. Figure \ref{test} shows a comparison of the solutions of the modifed model \ref{50} with the observed values. The MSE of predicted values
compared to the observed values for initializations after every \SI{40}{s} is \SI{0.66}{K^2}. When observations are provided after every \SI{60}{s} interval the MSE is found to be \SI{0.65}{K^2}. In case if the nominal model of heat equation is solved, the MSE is found to be \SI{12.23}{K^2} and \SI{14.49}{K^2} for observations provided every \SI{40}{s} and \SI{60}{s} respectively. In conclusion, the modified equation gives a better prediction accuracy than the nominal model for the second partition of the dataset which was untouched in the analysis done to modify the nominal equations.
\section{Summary and Discussion}
The main idea introduced in this article is briefly summarized as follows: The present work builds around the idea of finding "missing" terms in the evolution rule of a dynamical system by constraining the system to take values on the manifold formed by its observations. The information about the observed state of a system thus provides additional information of the forces acting on it. When applied to historical dataset of observations, we are able to generate examples of these extra terms for different instances of system states and time. These examples of extra terms (or forces) are then used to modify the evolution rule of the system to make improved predictions. In this process, additional terms estimated from system observations appear in the system equations while also retaining the known physics-based mechanistic terms.\par
In one of the examples this approach is applied to satellite orbit prediction. Satellites revisit the same sector of space after every revolution. This allows us to have many instances of extra force terms for a particular sector. The physical nature of the problem also suggests that force field in the same sector of space would not vary much in short span of time. So the previous values of extra forces can be used to modify the nominal gravitational model of the satellite dynamics to achieve reasonable accuracy. Historical precise orbit product from IGS is used to generate examples of extra terms in the satellite force model with respect to its position in space. With the added extra terms, the predicted position of the satellite is \SI{63.759}{m} away from the precise orbit position at the end of 2 hour prediction interval. The nominal gravitational model without any extra constraint force terms predicts the position approximately 8.5 times farther away from the precise position for the same prediction interval. When all the mechanistic terms are accounted, the force model for an orbiting satellite is computationally expensive. We started with a basic gravitational model. This improvement is achieved at a low computational cost for the prediction step than adding all other perturbation terms in the gravitational model. Also no extra data of other celestial bodies is needed. We expect the prediction accuracy to increase further if starting nominal model is more sophisticated.\par 
In another example, temperature is predicted for a metal rod undergoing cooling through one-dimensional heat conduction. Since the dataset used in this problem comes from one single physical experiment comprising of temperature measurement during conduction, temperatures attained are different during the analysis and prediction intervals. As such, the extra term in the heat equation has to be parametrized during the prediction step using the generated examples. Improved temperature prediction accuracy is achieved as compared to the nominal heat equation model. This example presents a different flavour of the same method in terms of application. In conclusion, in this study we have presented a method for dynamical system prediction by combining physics-based model with historical data. Numerical examples of satellite orbit prediction and temperature prediction in heat conduction experiment using this approach demonstrate improvement in predictions over the nominal physical models of these phenomena.
\section{Acknowledgement}
This work was partially supported by Indian Space Research Organisation through the RESPOND grant RES/ISRO-IISc JP/17-18 dated 4.12.2017.
\printbibliography
\end{document}